\documentclass[a4paper,11pt]{article}
\usepackage{jcappub}
\usepackage{graphicx}
\usepackage{amsmath}
\usepackage{amssymb}
\usepackage{amsfonts}
\usepackage{bm}

\def\be{\begin{align}}
\def\ee{\end{align}}
\def\bea{\begin{eqnarray}}
\def\eea{\end{eqnarray}}

\newcommand{\mc}[1]{\mathcal{#1}}
\newcommand{\f}[2]{\frac{#1}{#2}}

\title{\boldmath Cosmology of a Lorentz violating Galileon theory}
\author[a]{Zahra Haghani}
\author[b]{, Tiberiu Harko}
\author[c]{, Hamid Reza Sepangi}
\author[a]{and Shahab Shahidi}
\affiliation[a]{School of Physics, Damghan University, Damghan, 41167-36716,
Iran.}
\affiliation[b]{Department of Mathematics, University College London, Gower
Street,
London WC1E 6BT, United Kingdom.}
\affiliation[c]{Department of Physics, Shahid Beheshti University, G. C., Evin,
Tehran
19839, Iran.}

\emailAdd{z.haghani@du.ac.ir}
\emailAdd{t.harko@ucl.ac.uk}
\emailAdd{hr-sepangi@sbu.ac.ir}
\emailAdd{s.shahidi@du.ac.ir}

\date{\today }

\abstract{
We modify the scalar Einstein-aether theory by breaking the Lorentz
invariance of a gravitational theory
coupled to a Galileon type scalar field. This is done by introducing a Lagrange
multiplier term into the action, thus ensuring that the gradient of the scalar
field
is time-like, with unit norm.
The theory can also be considered as an extension to the mimetic dark matter
theory, by adding some derivative self interactions to the action, which keeps
the equation of motion at most second order in time derivatives.
The cosmological implications of the model are
 discussed in detail. In particular, for pressure-less baryonic matter, we show
that the
universe experiences a late time acceleration.
The cosmological implications of a special coupling between the
scalar field and the trace of the energy-momentum tensor are also explored.
}
\begin{document}
\maketitle
\flushbottom
\section{Introduction}
Lorentz invariance seems to be a fundamental property of nature which
has been tested within the context of the standard model of particle physics
with a
very high accuracy \cite{1}. However, in the gravity sector,  Lorentz
invariance cannot be tested to such a high precision. All the gravitational
experiments put a mild bound on the validity of Lorentz invariance
\cite{2}. In the dark sector of the universe, we have some constraints on
the breaking of  Lorentz invariance in  dark matter
\cite{darkmatter} and dark energy sectors \cite{darkenergy}. For a very
good review, see \cite{revdark}. This means that one can build a Lorentz
invariant theory which breaks the Lorentz invariance but satisfyie the experimental data. In this sense, it would be
plausible to
have a gravitational theory, describing the large scale structure and dynamics
of the universe which breaks Lorentz invariance.

Another motivation for considering
theories that break  Lorentz invariance comes from quantum
gravity. A very interesting example of such a theory is the Horava-Lifshitz
gravity where adding higher order Lorentz violating terms to the action
together with different scaling dimensions for space and time makes the theory
power counting renormalizable at the ultra-violet level \cite{3}. In the
infrared region, in the Horava-Lifshitz theory, the scaling
dimension of time becomes unity. Therefore, the Horava-Lifshitz theory
can in principle be written in a generally covariant form \cite{covHL}. However, due to a
fixed foliation of space-time into space-like hypersurfaces, the theory should
break Lorentz invariance dynamically.

One of the best candidates for a generally covariant gravitational theory with a preferred time-like direction is the Einstein-aether theory, proposed in
\cite{4}. The preferred direction can be imposed to the theory by introducing a time-like vector field to the action through a Lagrange multiplier. The action of the theory then becomes
\begin{align}  \label{act}
S_{ae}=\frac{1}{2\kappa^2}\int d^4x\sqrt{-g}\Big[R+K_{\lambda \sigma }^{\mu
\nu }\nabla _{\mu } u^{\lambda }\nabla _{\nu } u^{\sigma }+\lambda
\left(u_{\mu } u^{\mu }+1\right)\Big]+S_m,
\end{align}
where $\lambda $ is a Lagrange multiplier and the tensor $K^{\mu \nu}
_{\lambda \sigma}$ is
defined as
\begin{align}  \label{coef}
K^{\mu \nu}_{\lambda \sigma } = c_0g^{\mu \nu }g_{\lambda \sigma } +
c_1\delta ^{\mu } _{\sigma } \delta ^{\nu } _{\lambda }+ c_2\delta ^{\mu
}_{\lambda } \delta ^{\nu }_{\sigma } + c_3u^{\mu }u^{\nu } g_{\lambda
\sigma },
\end{align}
where $c_i$, $i=0,1,2,3$ are the dimensionless free parameters \cite{4}.
The stability of the theory has
also been studied and it turns out that the theory has a non-empty region in the
parameter space in which no instabilities occur \cite{7}. The PPN parameters
were
also obtained \cite{8} and it was shown that all PPN parameters are identical
to the standard GR if one imposes the conditions
$c_2=(-2c_1^2-c_1c_3+c_3^2)/3c_1$
and $c_4=-c_3^2/c_1$ respectively, for a discussion of this issue see also
\cite{9}.

The other interesting fact about the Einstein-aether theory is that it can be
considered as a covariant version of the Horava-Lifshitz theory in the infrared
sector.
In fact, one can prove
that the above theory is identical to the non-projectable Horava-Lifshitz
theory for a special case of the aether field, $u_\mu=N\nabla_\mu T$, where $N$
is
a normalization factor \cite{10}. However, the action \eqref{act} can only
produce the lowest terms of the Horava-Lifshitz theory and the search for a
generally covariant theory, which produces higher order terms of the
Horava-Lifshitz theory is still a subject of research.

It will also be interesting if one could write an
action for a scalar field with a preferred time-like direction. In
this case the covariant derivative of the scalar field can play the role of a
time-like vector. In
\cite{SEA} the authors have suggested such a gravitational model by
substituting
$%
u_\mu$ by $\nabla_\mu\phi$ in the action \eqref{act}. However, this choice
separates only the longitudinal mode of the vector field and also
implies that $\nabla_\mu\phi$ has a unit norm. It was shown in
\cite{jacob2} that the theory proposed in \cite{SEA} is identical to the
projectable
Horava-Lifshitz gravity  which has instabilities \cite{ins,HL}. The problem lies
in
the aether
action itself, where substituting the aether field with $\nabla_\mu\phi$
produces some fourth order derivative terms. These higher order derivative terms
will produce ghost instabilities to the action. Hence, in order to write a
scalar action which breaks  the Lorentz invariance dynamically, one should
introduce a kinetic term for the scalar field which has no instabilities.

One of the main motivations of the present paper is to write down an action for
a ``healthy'' scalar-aether theory. We first note that in writing the
action \eqref{act}, the authors have introduced the
most general canonical kinetic term for a vector field. In the scalar case, the canonical kinetic term is of the form $\partial_\mu\phi\partial^\mu\phi$, which
is the
same as the term used to break the Lorentz invariance. In fact, one
adds a term $\lambda(\partial_\mu\phi\partial^\mu\phi+1)$ to the action
where $\lambda$ is a Lagrange multiplier.
In this paper, we want to add some higher order kinetic interactions to the theory in a way that the resulting theory has no Ostrogradski instability. In this sense one should add some healthy higher order derivative interactions to the action. Such terms are known as Galileons
\cite{gali}.

The Galileons
are scalar fields where  having even higher than second order time derivatives
in the action produce second order field equations. The Galileons were
originally written in  flat Minkowski space \cite{gali} where they have an
additional symmetry
\begin{align}  \label{i2}
\phi\rightarrow\phi+b_\mu x^\mu+c,
\end{align}
which is lost in the covariant version \cite{covgali} \footnote{The most general theory of two graviton degrees of freedom (dof) plus a scalar dof writing in terms of curvature invariants and the covariant derivatives of the scalar field was first introduced by Horndeski
in \cite{horn}, and independently rediscovered in \cite{reHor}. The covariant Galileon theory which we are using in this paper is then a special case of Horndeski theory.}.
In order to have a covariant a Galileon theory, one can replace the partial derivatives with the covariant derivatives. However, because of the
non-commutativity
of the covariant derivatives, one obtains some higher order terms which produce Ostrogradski
instabilities in the curved background. In order to remove these
non-healthy terms one should add to the action some new terms which do not
respect
the Galilean symmetry \eqref{i2}. This will produce a covariant Galileon action \cite{covgali}. One can also generalize the Galileons by
substituting the coefficients of each term with an arbitrary function of
the scalar field \cite{11}.

Recently, a very interesting model for  dark matter sector of the universe
was proposed in the literature, known as the ``mimetic dark matter'' \cite{14}.
The
model has originally suggested a new action for a conformally
invariant gravitational theory. In this sense, the authors have added a scalar
field
to the theory and built the gravitational action using the effective metric
\begin{align}\label{i3}
g_{\mu\nu}=(\bar{g}^{\alpha\beta}\partial_\alpha\phi\partial_\beta\phi)\bar{g}_{
\mu\nu}.
\end{align}
The theory is obviously invariant under the change
$\bar{g}_{\mu\nu}\rightarrow\Omega^2\bar{g}_{\mu\nu}$, with $\Omega$ an
arbitrary function.
One can easily observe that in this theory the scalar
field $\phi$ is subject to the constraint
$g^{\mu\nu}\partial_\mu\phi\partial_\nu\phi=-1$, which implies that the vector
$u_\mu=\nabla_\mu\phi$ is time-like \cite{14,15}. So the vector field $u_\mu$ will be tangent to time-like geodesics, and will correspond to the 4-velocity of a dust particle. This geometrical dust corresponds to dark matter. The mimetic theory can then explain the dark matter content of the universe as a consequence of the constraint equation $g^{\mu\nu}\partial_\mu\phi\partial_\nu\phi=-1$. This also happens in the
scalar Einstein-aether theory where one adds a similar constraint to the
action through a Lagrange multiplier. This suggests that the  model introduced
in the present paper can
be considered as an extension of the mimetic dark matter theory with the
addition of the
most general higher derivative self interaction which limits the equations of
motion to second order. Such a modification makes dark matter sector of the
universe imperfect \cite{17}. We should note that one can generalize the constraint equation by rewriting it as $g^{\mu\nu}\partial_\mu\phi\partial_\nu\phi=-\mu(\phi)^2$ \cite{cauchy}. This condition implies that the vector field $u_\mu=\partial_\mu\phi/\mu$ is tangent to time-like geodesics, corresponding to the 4-velocity of dark matter.

The present paper is organized as follows. In Section~\ref{sec1} we introduce
the action
of the Lorentz violating Galileon theory and  obtain the equations of motion. In
Section~\ref{sect3} we study the cosmological implications of the theory,
showing that it can explain the accelerated expansion of a matter dominated
universe, having only zero pressure dust in the energy-momentum of the universe. In
Section \ref{sec4} we  add a special interaction between the aether field and the trace of the
energy momentum tensor of ordinary matter and study the cosmology of this
modified gravity model. Last Section is devoted a discussion of our results.

\section{Generalized scalar Einstein-aether theory}\label{sec1}

We propose a generalization of the scalar Einstein-aether theory based on a
gravitational action of the form
\begin{align}  \label{1}
S=\int d^4x\sqrt{-g}\bigg[\kappa^2 R+\alpha_3\mathcal{L}_3+\alpha_4\mathcal{L%
}_4+\alpha_5\mathcal{L} _5+\lambda(\phi_\mu\phi^\mu+1)+\mathcal{L}_m\bigg],
\end{align}
where $\mathcal{L}_m$ is the matter Lagrangian and $\phi_\mu\equiv\nabla_%
\mu\phi$. In the above action we have dropped the tadpole Galileon term and
also absorbed the quadratic term into the term $\lambda(\phi_\mu\phi^\mu+1)$.  We have also
introduced the terms $\mathcal{L}_i$, $i=1,2,3$, defined as \cite{covgali}
\begin{align}  \label{2}
\mathcal{L}_3&=(\phi_\alpha\phi^\alpha)\Box\phi, \\
\mathcal{L}_4&=(\phi_\alpha\phi^\alpha)\bigg[2(\Box\phi)^2-2\phi_{\mu\nu}%
\phi^{ \mu\nu} -\frac{1}{2}(\phi_\mu\phi^\mu)R\bigg], \\
\mathcal{L}_5&=(\phi_\alpha\phi^\alpha)\bigg[(\Box\phi)^3-3(\phi_{\mu\nu}%
\phi^{ \mu\nu}
)\Box\phi+2\phi_\mu^{~~\nu}\phi_\nu^{~~\rho}\phi_\rho^{~~\mu}-6\phi_\mu%
\phi^{ \mu\nu}G_{\nu\rho}\phi^\rho\bigg].
\end{align}
The action \eqref{1} is similar to the action for mimetic dark matter with
derivative interaction terms for a scalar field \cite{15}.

Variation of the action with respect to $\lambda$ gives
\begin{align}  \label{3}
\phi_\mu\phi^\mu=-1,
\end{align}
which, as was mentioned before, defines a preferred direction for the
space-time, and dynamically breaks the Lorentz symmetry of the theory. After
using Eq.~\eqref{3} and its derivative
\begin{align}  \label{3.1}
\phi^\mu\phi_{\mu\nu}=0,
\end{align}
we obtain the reduced form of the metric field equations as
\begin{align}  \label{4}
\kappa^2 G_{\mu\nu}=T_{\mu\nu}+\alpha_3 T^3_{\mu\nu}+\alpha_4
T^4_{\mu\nu}+\alpha_5 T^5_{\mu\nu}-\lambda\phi_\mu\phi_\nu,
\end{align}
where $T_{\mu\nu}$ is the ordinary matter energy momentum tensor defined as
$$T_{\mu\nu}=\f{-2}{\sqrt{-g}}\f{\delta (\sqrt{-g}\mathcal{L}_m)}{\delta 
g^{\mu\nu}},$$
and we 
have defined
\begin{align}  \label{5}
T^3_{\mu\nu}&=-\phi_\mu\phi_\nu\Box\phi, \\
T^4_{\mu\nu}&=-2\phi_{\mu\nu}\Box\phi+2\phi_{\mu\rho}\phi^{\rho}_{~~\nu}+%
\big((\Box\phi)^2 -\phi_{\rho\sigma}\phi^{\rho\sigma}\big)(g_{\mu\nu}
-2\phi_\mu\phi_\nu)  \notag \\
&\qquad -\phi_\mu\phi_\nu
R+2\phi^\rho(R_{\rho\mu}\phi_\nu+R_{\rho\nu}\phi_\mu)+\frac{1}{2}G_{\mu\nu}
-2\phi_\rho R^{\rho\sigma}\phi_\sigma g_{\mu\nu}+2\phi^\rho\phi^\sigma
R_{\mu\rho\nu\sigma},
\end{align}
and
\begin{align}  \label{6}
T^5_{\mu\nu}&=-3\big((\Box\phi)^2-\phi_{\sigma\lambda}\phi^{ \sigma\lambda}%
\big)\phi_{\mu\nu}+\Box\phi\big(6\phi_{\mu\sigma} \phi^\sigma_{~~\nu}-\frac{3%
}{2}\phi_\mu\phi_\nu R\big)+3\phi^\sigma\Box\phi\big(R_{\sigma\mu}\phi_%
\nu+R_{\sigma\nu} \phi_\mu\big)  \notag \\
& +3\Box\phi\phi^\sigma\phi^\lambda
R_{\mu\sigma\nu\lambda}-6\phi_{\mu\sigma}\phi^{\sigma\rho}\phi_{\rho\nu}+3%
\phi_{\sigma\lambda}R^{\sigma\lambda}\phi_\mu\phi_\nu-3\phi_\sigma
R^{\sigma\lambda}\big(\phi_{\lambda\mu}\phi_\nu+\phi_{\lambda\nu} \phi_\mu%
\big)  \notag \\
&-3\phi^\sigma\phi^{\lambda\rho}\big(R_{\mu\lambda\sigma\rho}
\phi_\nu+R_{\nu\lambda\sigma\rho}\phi_\mu\big)+3 \phi^\sigma\phi^\lambda\big(%
R_{\mu\sigma\lambda\rho}\phi^\rho_{~~\nu}+R_{
\nu\sigma\lambda\rho}\phi^\rho_{~~\mu}\big)+3
\phi_\sigma\phi_\lambda\phi_{\rho\kappa}R^{\sigma\rho\lambda\kappa}g_{\mu\nu}
\notag \\
&+\bigg((\Box\phi)^3-3\Box\phi(\phi_{\rho\sigma}\phi^{\rho\sigma})
+2\phi_{\rho\sigma}\phi^{\sigma\lambda} \phi_\lambda^{~~\rho}\bigg)%
(g_{\mu\nu}-\phi_{\mu}\phi_{\nu})+3(\phi_\sigma
R^{\sigma\lambda}\phi_\lambda)(\phi_{\mu\nu}-\Box\phi g_{\mu\nu}).
\end{align}
The aether field equation of motion can be written as
\begin{align}  \label{7}
\alpha_3\mathcal{E}_3+\alpha_4\mathcal{E}_4+\alpha_5\mathcal{E}
_5-2\nabla_\mu\big(\lambda\phi^\mu\big)=0,
\end{align}
where we have defined
\begin{align}  \label{8}
\mathcal{E}_3&=-2(\Box\phi)^2+2R_{\mu\nu}\phi^\mu\phi^\nu+2\phi_{\mu\nu}
\phi^{\mu\nu}, \\
\mathcal{E}_4&=-4(\Box\phi)^3-8\phi_{\mu\nu}\phi^{\nu\sigma}\phi_\sigma^{~~%
\mu} +12\Box\phi(\phi_{\mu\nu}\phi^{\mu\nu})-2(\Box\phi)R  \notag \\
&\qquad+8(\Box\phi)\phi_\mu
R^{\mu\nu}\phi_\nu+4\phi_{\mu\nu}R^{\mu\nu}-8\phi_\mu\phi_\nu\phi_{\sigma%
\rho}R^ {\mu\rho\nu\sigma},
\end{align}
and
\begin{align}  \label{9}
\mathcal{E}_5=&-2(\Box\phi)^4+3(\Box\phi)^2\bigg(4\phi_{\mu\nu}\phi^{
\mu\nu}-R+2\phi_\mu R^{\mu\nu}\phi_\nu\bigg)-16\Box\phi(\phi_{\mu\nu}\phi^{%
\nu\rho}\phi_\rho^{~~\mu}) +12(\Box\phi)\phi_{\mu\nu}R^{\mu\nu}  \notag \\
&-12(\Box\phi)\phi_\mu\phi_\nu\phi_ {
\rho\sigma}R^{\mu\rho\nu\sigma}-6(\phi_{\mu\nu}\phi^{\mu\nu})(\phi_{\rho%
\sigma}
\phi^{\rho\sigma})+12\phi_{\mu\nu}\phi^{\nu\rho}\phi_{\rho\sigma}\phi^{%
\sigma\mu }+3(\phi_{\mu\nu}\phi^{\mu\nu})R  \notag \\
&-6(\phi_{\mu\nu}\phi^{\mu\nu})(\phi_\rho
R^{\rho\sigma}\phi_\sigma)-12\phi_{\nu\rho}R^{\rho\sigma}\phi_\sigma^{~~\nu}
-6\phi_{\nu\rho}\phi_{\sigma\lambda}R^{\nu\sigma\rho\lambda}
+12\phi_\mu\phi_\nu\phi_{\rho\sigma}\phi^\sigma_{~~\lambda}R^{\mu\rho\nu%
\lambda}  \notag \\
&+3(\phi_\nu R^{\nu\rho}\phi_\rho)R-6\phi_\nu
R^{\nu\rho}R_{\rho\sigma}\phi^\sigma-6\phi_\nu\phi_\rho
R_{\sigma\lambda}R^{\nu\sigma\rho\lambda}+3\phi_\nu\phi_\rho
R^\nu_{~~\sigma\kappa\lambda}R^{\rho\sigma\kappa\lambda}.
\end{align}

Note that the energy-momentum tensors $T^i_{\mu\nu}$ have the property that
their covariant derivatives become proportional to $\mc{E}_i$ \cite{gali}
\begin{align}\label{9.1}
\nabla^\mu T^i_{\mu\nu}=\f{1}{2}\phi_\nu\mc{E}_i, \qquad i=1,2,3.
\end{align}
So, the covariant derivative of the equation \eqref{4} is reduced to
\begin{align}\label{9.2}
\nabla^\mu T_{\mu\nu}=-\f12\phi_\nu
\big[\alpha_3\mc{E}_3+\alpha_4\mc{E}_4+\alpha_5\mc{E}
_5-2\nabla_\mu(\lambda\phi^\mu)\big].
\end{align}
The energy-momentum tensor of ordinary matter is then conserved if the
equation of motion for the aether field is satisfied. In the following we will assume that the energy
momentum tensor of ordinary matter is conserved $\nabla^\mu T_{\mu\nu}=0$.

\section{Cosmological implications}

\label{sect3}

In this Section we study the cosmological implications of the Lorentz violating
Galileon theory. We
will restrict our study to homogeneous and isotropic cosmological models
with the line element given by the flat Friedmann-Robertson-Walker metric
\begin{align}  \label{cos-1}
ds^2=-dt^2+a^2(t)d\vec{x}^2.
\end{align}
We also assume that the aether field and the Lagrange multiplier are
homogeneous and therefore have  the form $\phi=\phi(t)$ and $%
\lambda=\lambda(t) $, respectively. We also assume that the matter
energy-momentum tensor, describing the matter content of the universe,
has a perfect fluid form
\begin{align}  \label{cos-1.5}
T^\mu_{~~\nu}=\mathrm{diag}\big[-\rho(t),p(t),p(t),p(t)\big],
\end{align}
where $\rho (t)$ is the matter energy density, while $p(t)$ represents the
thermodynamic pressure.

\subsection{The cosmological evolution equations}

Due to the choice of the geometry as homogeneous and isotropic, the
constraint equation Eq.~\eqref{3} completely determines the aether field $%
\phi$ as
\begin{align}  \label{cos-2}
\phi=t+c_1,
\end{align}
where $c_1$ is an arbitrary integration constant. For a homogeneous and
isotropic cosmological model, Eq.~\eqref{cos-2} is obvious since the FRW
metric already has a time-like preferred direction $\partial/\partial t$,
and the aether vector should be identical to that direction up to a shift.

The metric and scalar field equations can then be obtained from Eqs.~%
\eqref{4} and \eqref{7} as
\begin{align}
&\frac{3}{2}(15\alpha_4+2\kappa^2)H^2-21\alpha_5 H^3-3\alpha_3
H-\rho+\lambda=0,  \label{9.1} \\
&(3\alpha_4+2\kappa^2-6\alpha_5H)\dot{H}+\frac{3}{2} (3\alpha_4+2%
\kappa^2)H^2-6\alpha_5H^3+p=0 ,  \label{9.2}
\end{align}
and
\begin{align}  \label{cos-5}
6(15\alpha_5H^2-12\alpha_4H+\alpha_3)\dot{H} +90\alpha_5H^4-108%
\alpha_4H^3+18\alpha_3H^2-2(3\lambda H+\dot{\lambda})=0.
\end{align}
In Eqs.~(\ref{9.1}) - (\ref{cos-5}) $H=H(t)=\dot{a}(t)/a(t)$ denotes the
Hubble parameter.

One can solve the aether equation for the Lagrange multiplier to obtain
\begin{align}  \label{10}
\lambda=3H(\alpha_3-6\alpha_4H+5\alpha_5H^2)+\frac{c_2}{a^3},
\end{align}
where $c_2$ is an arbitrary constant of integration. Substituting Eq.~%
\eqref{10} into Eqs.~\eqref{9.1} and \eqref{9.2}, one obtains
\begin{align}
&\frac{3}{2}(2\kappa^2+3\alpha_4)H^2-6\alpha_5H^3+\frac{c_2}{a^3}-\rho=0,
\label{11} \\
&(3\alpha_4+2\kappa^2-6\alpha_5H)\dot{H}+\frac{3}{2} (3\alpha_4+2%
\kappa^2)H^2-6\alpha_5H^3+p=0.  \label{12}
\end{align}

One can easily see that $\alpha_5$ introduces higher order Hubble parameter
to the Friedmann equation and $\alpha_4$ modifies the gravitational
constant of the theory. Solving Eq.~\eqref{11} for $\rho$ gives
\begin{align}  \label{13}
\rho=\frac{3}{2 } H^2\big(3\alpha_4+2\kappa^2-4\alpha_5 H\big)+\frac{c_2}{a^3%
}.
\end{align}

One of the most common equations of state, which has been used extensively
to study the properties of compact objects at high densities is the
linear barotropic equation of state, with $p = \omega \rho $ with $\omega =
\mathrm{constant} \in [0,1]$. Assuming the equation of state of the ordinary
matter as having a linear barotropic form and substituting Eq.~\eqref{13}
to Eq.~\eqref{12}, we obtain the basic cosmological evolution equation of
our model as
\begin{align}  \label{14}
(2\kappa^2+3\alpha_4-6\alpha_5H)\dot{H}
+\f32(\omega+1)\left(2\kappa^2+3\alpha_4-4\alpha_5H\right)H^2+\f{\omega
c_2}{a^3}=0.
\end{align}

\subsection{The general dynamics of the SEA cosmological models}

In order to obtain a simpler form of Eq.~(\ref{14}) we rescale the
parameters $\alpha _{4}$, $\alpha _{5}$ and $c_{2}$ so that
\begin{align}
\alpha _{4}=\frac{2\kappa ^{2}}{3}m,\quad\alpha _{5}=\frac{\kappa ^{2}}{3}%
n_{1},\quad c_{2}=2\kappa ^{2}s_{1},
\end{align}%
where $m$, $n_{1}$, $s_{1}$ are constants. Then Eq.~(\ref{14}) takes the
form
\begin{align}
(1+m-n_{1}H)\dot{H}+\frac{3}{2}\left( \omega +1\right) \left[ 1+m-\frac{2}{3}%
n_{1}H\right] H^{2}+\frac{\omega s_{1}}{a^{3}}=0.  \label{301}
\end{align}%
We introduce as an independent variable the redshift $z$, defined as $1+z=1/a$.
Therefore
\begin{align}
\frac{dH}{dt}=\frac{dH}{dz}\frac{dz}{dt}=-(1+z)H\frac{dH}{dz}.
\end{align}

Moreover, we represent $H(z)$ as $H(z)=H_{0}h(z)$, where $H_{0}$ is the
present value of the Hubble parameter. By rescaling again the coefficients $%
n_{1}$ and $s_{1}$ so that
\begin{align}
n_{1}=\frac{n}{H_{0}},\quad s=\frac{s_{1}}{H_{0}^{2}},
\end{align}%
we obtain the basic evolution equation of the Hubble parameter as
\begin{align}
(1+z)h(z)\left[ 1+m-nh(z)\right] \frac{dh}{dz}=\frac{3}{2}\left( \omega
+1\right) \left[ 1+m-\frac{2}{3}nh(z)\right] h^{2}(z)+\omega s(1+z)^{3}.
\label{331}
\end{align}%
By rescaling the density as $\rho (z)=2\kappa ^{2}H_{0}^{2}r(z)$ one obtains
\begin{align}
r(z)=\frac{3}{2}\left[ 1+m-\frac{2}{3}nh(z)\right] h^{2}(z)+s(1+z)^{3}.
\end{align}
By rescaling the constant $\alpha _{3}$ as $\alpha _{3}=2\kappa ^{2}uH_{0}$,
where $u$ is a constant, and the Lagrange multiplier  $\lambda $ as $%
\lambda =2\kappa ^{2}H_{0}^{2}\Lambda $, we obtain

\begin{align}
\Lambda \left( z\right) =3h(z)\left[ u-2mh(z)+\frac{5}{6}nh^{2}(z)\right]
+s\left( 1+z\right) ^{3}.
\end{align}

As an indicator of the accelerated expansion we introduce the deceleration
parameter, defined as
\begin{align}
q=\frac{d}{dt}\frac{1}{H(t)}-1=-\frac{\dot{H(t)}}{H^{2}(t)}-1=(1+z)\frac{1}{H(z)
}\frac{dH(z)}{dz}-1.
\end{align}%
With the use of Eq.~(\ref{301}) it follows that the deceleration parameter
can be expressed as

\begin{align}
q=\frac{3}{2}\left( \omega +1\right) \frac{2\kappa ^{2}+3\alpha _{4}-4\alpha
_{5}H}{2\kappa ^{2}+3\alpha _{4}-6\alpha _{5}H}+\frac{\omega c_{2}}{%
a^{3}H^{2}\left( 2\kappa ^{2}+3\alpha _{4}-6\alpha _{5}H\right) }-1,
\end{align}%
or, equivalently

\begin{align}
q=\frac{3}{2}\left( \omega +1\right) \frac{1+m-(2n_{1}/3)H}{1+m-n_{1}H}+%
\frac{\omega s_{1}}{a^{3}H^{2}\left( 1+m-n_{1}H\right) }-1.
\end{align}
As a function of redshift the deceleration parameter is obtained as
\begin{align}
q(z)=\frac{3}{2}\left( \omega +1\right) \frac{1+m-\left( 2/3\right) nh(z)}{%
1+m-nh(z)}+\frac{\omega s\left( 1+z\right) ^{3}}{h^{2}(z)\left[ 1+m-nh(z)%
\right] }-1.
\end{align}%
The sign of the deceleration parameter indicates the nature of the
expansionary evolution. If $q>0$, the cosmological expansion is
decelerating, while negative values of $q$ indicate an accelerating dynamics.

In the following we consider the cosmological implications of the present
model for several equations of state of the cosmological matter. We first
investigate the high density phase of the evolution of the universe with
matter obeying the stiff and radiation equations of state. We then analyze
in detail the behavior of  dust (zero thermodynamic pressure)
cosmological models.

\subsection{The stiff and radiation fluid cases}

We begin our study of the cosmology of the Lorentz violating Galileon model
by analyzing the high density universe described by the Zeldovich
(stiff) equation of state and by the radiation equation of state,
respectively. The Zeldovich equation of state $p=\rho$ is valid for densities
significantly higher than nuclear densities $\rho _n$, $\rho > 10\rho _n$. It
can be
obtained by constructing a relativistic Lagrangian which allows bare nucleons
to interact attractively through scalar meson exchange, and repulsively through
the
exchange of a slightly more massive vector meson \cite{60}. In the
non-relativistic
limit both the quantum and classical theories lead to Yukawa-type
self-interaction potentials.
But at the highest possible matter densities the vector meson exchange
dominates. By using a
mean field approximation for the nuclear interactions, it follows that in the
extreme limit of very high
densities the thermodynamic pressure tends to the energy density, $p
\rightarrow
\rho $.
In this high density limit the  speed of sound $c_s$ tends to one,  $c_s
^2=dp/d\rho \rightarrow 1$. Therefore the
Zeldovich (stiff fluid) equation of state satisfies the causality condition
with the speed of sound less than or equal to the speed of light. The radiation fluid
satisfies an equation of state of the form $p=(1/3)\rho $.

\begin{figure}[h]
\hspace{-.3cm}
\begin{minipage}{0.5\textwidth}
 \centering
\includegraphics[width=7.4cm]{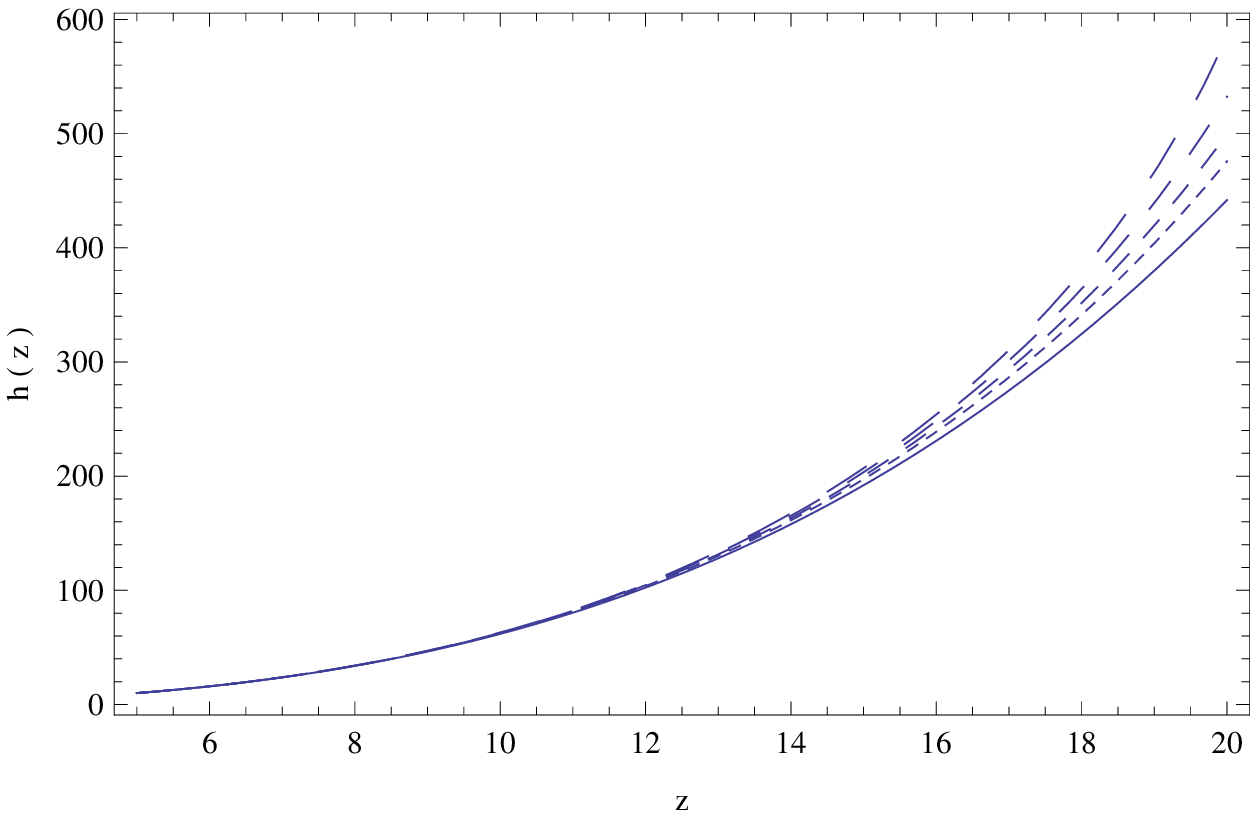}
\end{minipage}
\begin{minipage}{0.5\textwidth}
\centering
\includegraphics[width=7.7cm]{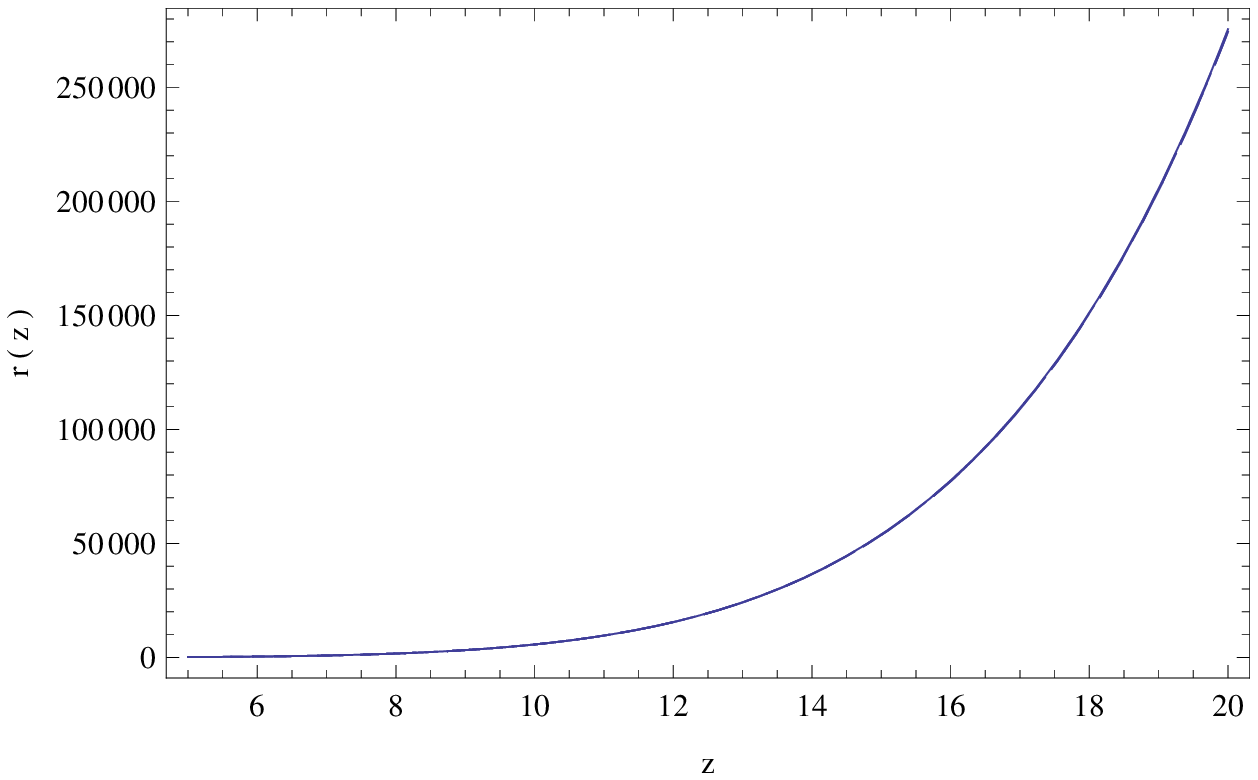}
\end{minipage}
\caption{Variation of the Hubble parameter $h(z)$ (left figure) and of the
energy density $r(z)$ (right figure) as a function of $z$ for a stiff fluid
filled universe, for different values of the parameters $m$, $n$, $s$: $m =
0.0001$, $n = 0.0002$, $s = 0.0003$ (solid curve), $m = 0.0004$, $n = 0.0006$%
, $s = 0.0008$ (dotted curve), $m = 0.0006$, $n = 0.0008$, $s = 0.001$
(short dashed curve), $m = 0.0008$, $n = 0.001$, $s = 0.0012$ (dashed
curve), and $m = 0.001$, $n = 0.0012$, $s = 0.0014$, respectively.}
\label{fig1}
\end{figure}

\begin{figure}[h]
\centering
\includegraphics[width=8cm]{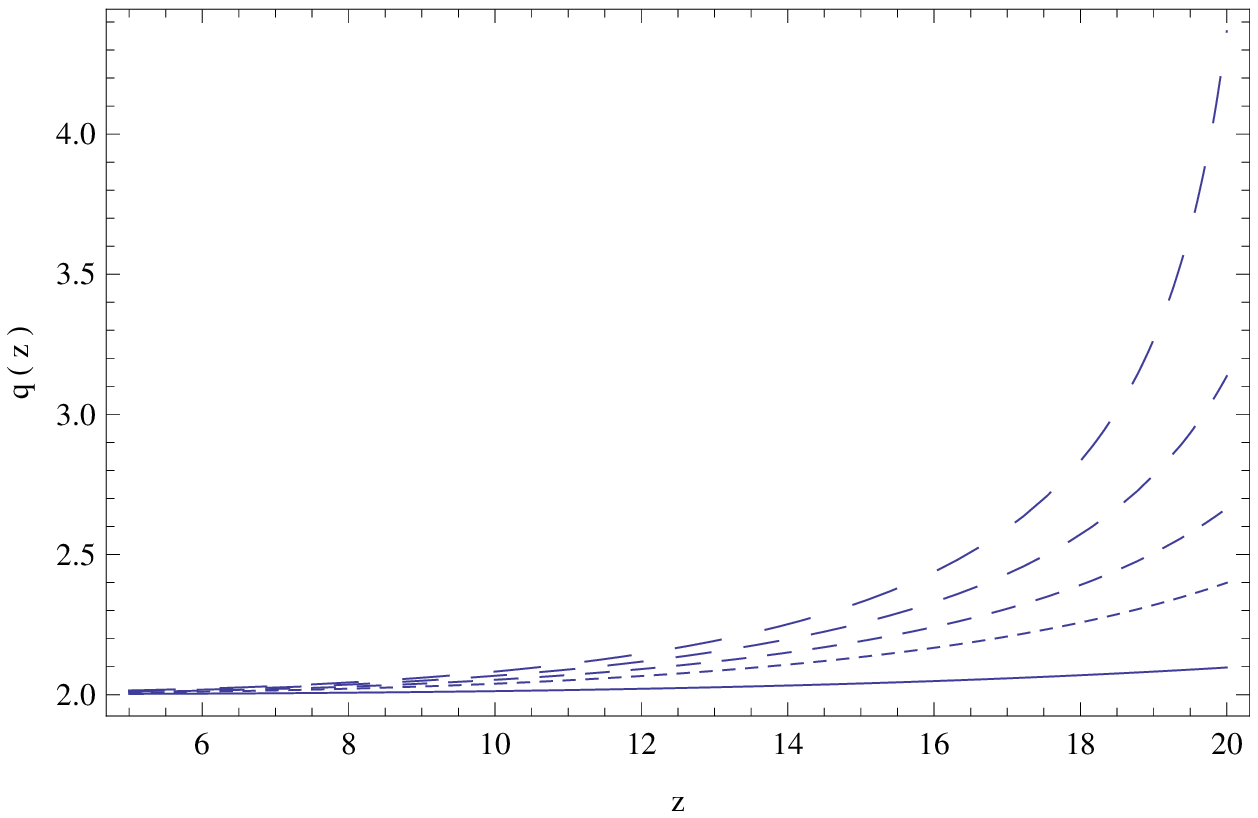}
\caption{Variation with respect to $z$ of the deceleration parameter $q(z)$
for a stiff fluid filled universe, for different values of the parameters $m$%
, $n$, $s$: $m = 0.0001$, $n = 0.0002$, $s = 0.0003$ (solid curve), $m =
0.0004$, $n = 0.0006$, $s = 0.0008$ (dotted curve), $m = 0.0006$, $n =
0.0008 $, $s = 0.001$ (short dashed curve), $m = 0.0008$, $n = 0.001$, $s =
0.0012$ (dashed curve), and $m = 0.001$, $n = 0.0012$, $s = 0.0014$,
respectively.}
\label{fig3}
\end{figure}

The variations of the Hubble parameter, energy density and deceleration
parameter are represented, as functions of the redshift $z$, and for
different values of the parameters $m$, $n$ and $s$ for a stiff fluid filled
universe, in Figs.~\ref{fig1} and \ref{fig3}, while the variation of the same
quantities for a radiation filled universe are represented in Figs.~\ref%
{fig4} and \ref{fig6}, respectively.

\begin{figure}[h]
\begin{minipage}{0.5\textwidth}
\includegraphics[width=7.5cm]{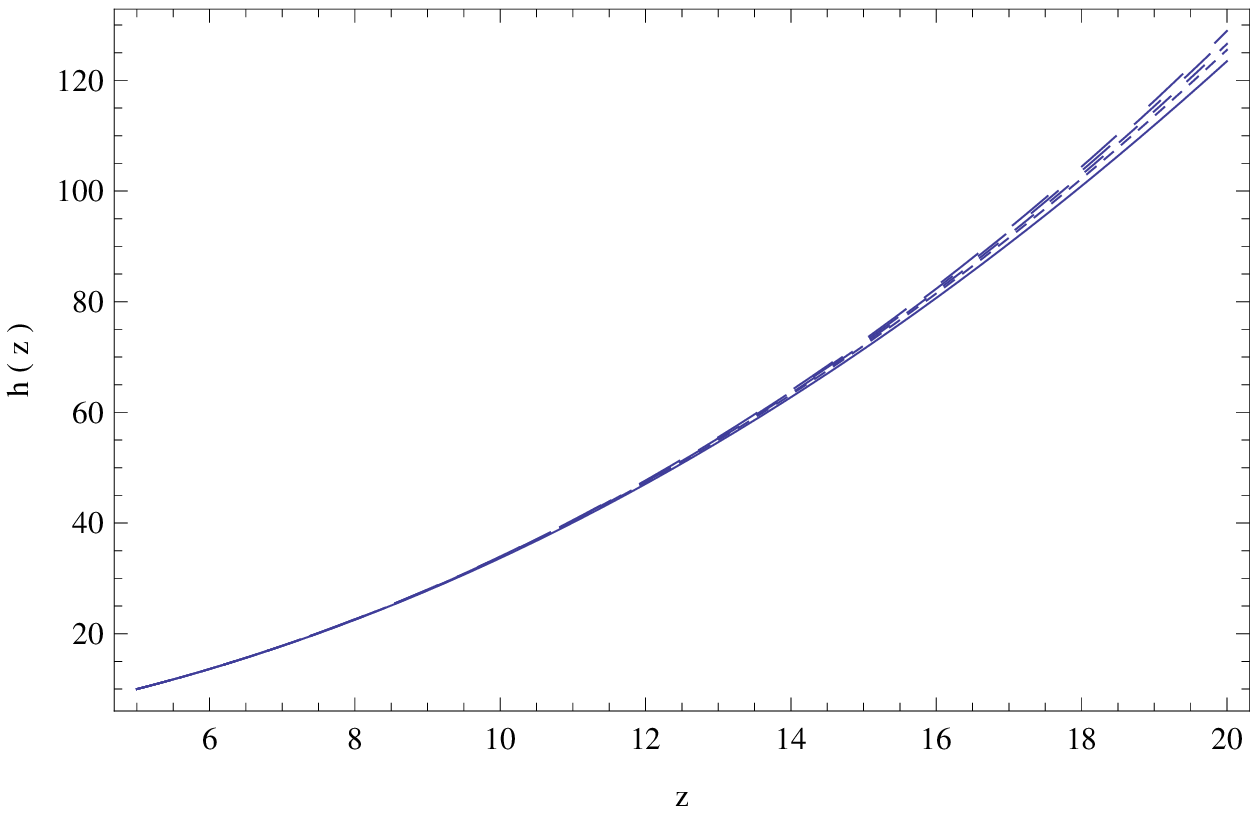}
\end{minipage}
\begin{minipage}{0.5\textwidth}
\includegraphics[width=7.5cm]{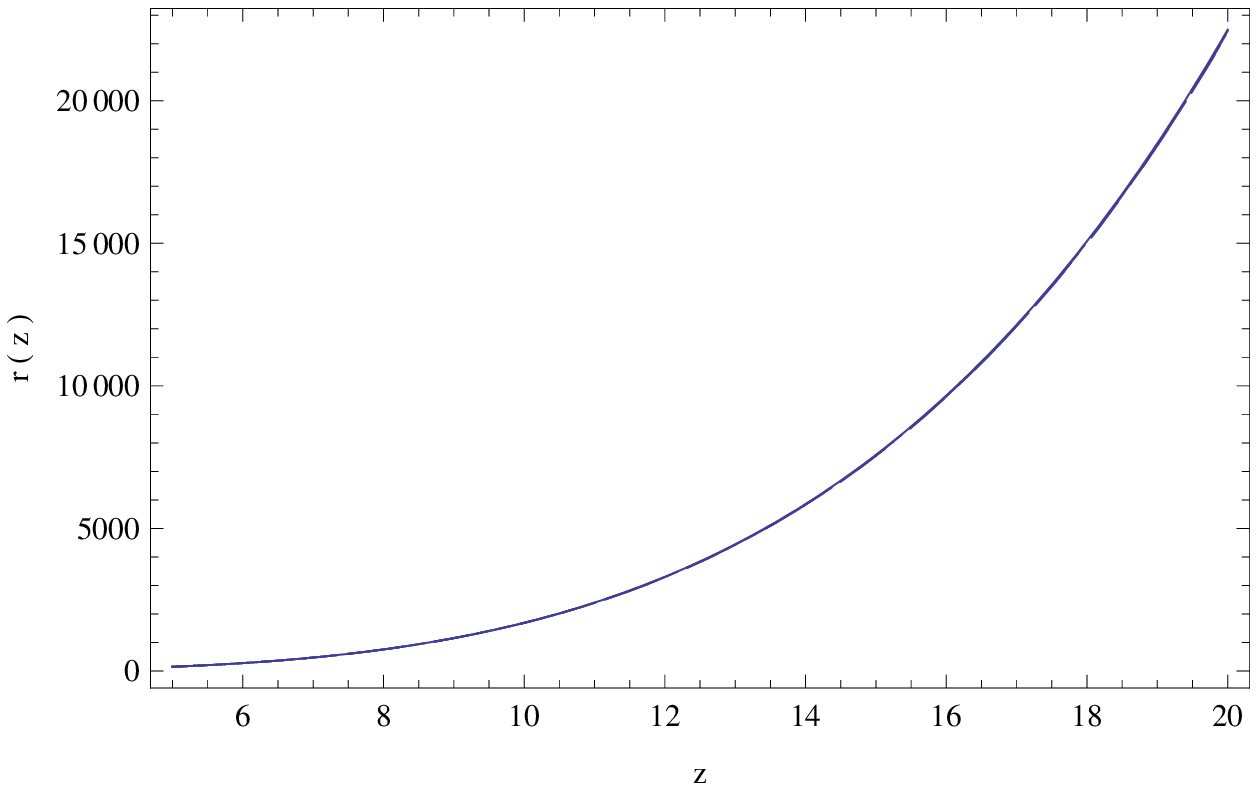}
\end{minipage}
\caption{Variation as a function of $z$ of the Hubble parameter $h(z)$ (left
figure) and of the matter energy density $r(z)$ (right figure) for a
radiation fluid filled universe, for different values of the parameters $m$,
$n$, $s$: $m = 0.0001$, $n = 0.0002$, $s = 0.0003$ (solid curve), $m =
0.0004 $, $n = 0.0006$, $s = 0.0008$ (dotted curve), $m = 0.0006$, $n =
0.0008$, $s = 0.01$ (short dashed curve), $m = 0.0008$, $n = 0.01$, $s =
0.012$ (dashed curve), and $m = 0.001$, $n = 0.012$, $s = 0.014$,
respectively.}
\label{fig4}
\end{figure}

\begin{figure}[h]
\centering
\includegraphics[width=8cm]{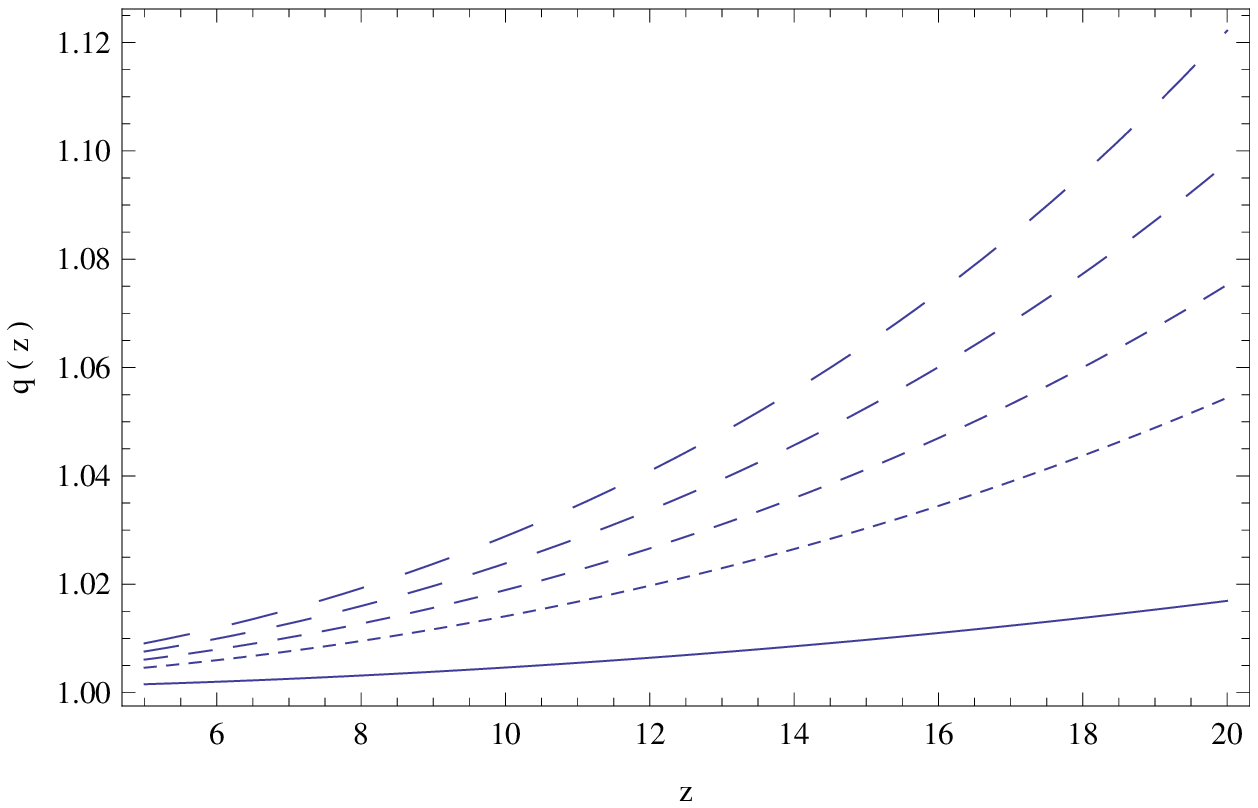}
\caption{Variation with respect to $z$ of the deceleration parameter $q(z)$
for a radiation fluid filled universe, for different values of the
parameters $m$, $n$, $s$: $m = 0.0001$, $n = 0.0002$, $s = 0.0003$ (solid
curve), $m = 0.0004$, $n = 0.0006$, $s = 0.0008$ (dotted curve), $m = 0.0006$%
, $n = 0.0008$, $s = 0.01$ (short dashed curve), $m = 0.0008$, $n = 0.01$, $%
s = 0.012$ (dashed curve), and $m = 0.001$, $n = 0.012$, $s = 0.014$,
respectively.}
\label{fig6}
\end{figure}

In both the stiff and radiation fluid cases we have studied the evolution of
the high density universe in the redshift range $5\leq z\leq 20$, with the
initial condition $h(5)=10$. The behavior of the cosmological models for
both equation of states is similar. For the range of considered parameters $%
m $, $n$ and $s$, the Hubble function is a monotonically increasing function
of $z$ (monotonically decreasing in time), as is the energy density. The
evolution of the deceleration parameter shows a decelerating expansion, with
values of the order of $q(5)\approx 2$ in the case of the stiff fluid
universe and with $q(5)\approx 1$ in the case of the radiation fluid. Of
course, modifying the numerical values of $m$, $n$ and $s$ may lead to
significantly different values of the cosmological parameters.

\subsection{The universe filled with dust}

For a universe filled with dust, $\omega =0$,  the time
evolution equation Eq.~(\ref{301}) for the Hubble parameter takes the form

\begin{align}
(1+m-n_{1}H)\dot{H}+\frac{3}{2}\left[ 1+m-\frac{2}{3}n_{1}H\right] H^{2}=0.
\label{eq_H}
\end{align}

Eq. (\ref{eq_H}) admits a de Sitter type solution $H=H_{0}={\rm constant}$,
corresponding to
\begin{align}
H_{0}=\frac{3(1+m)}{2n_{1}}.
\end{align}

For this choice of parameters the expansion is exponential with the
scale factor given by $a=a_{0}e^{H_{0}t}$, with the deceleration parameter
having the value $q=-1$. During a de Sitter type phase the energy density
of the universe is given by $\rho =c_{2}/a^{3}$, and it tends exponentially
to zero. For arbitrary values of the parameters the general solution of Eq.~(%
\ref{eq_H}) is given by
\begin{align}
\frac{3}{2}\left( t-t_{0}\right) =\frac{1}{H}+\frac{n_{1}}{3(1+m)}\ln \frac{H%
}{3+2m-2n_{1}H}.
\end{align}

\begin{figure}[h]
\begin{minipage}{0.5\textwidth}
\includegraphics[width=7.5cm]{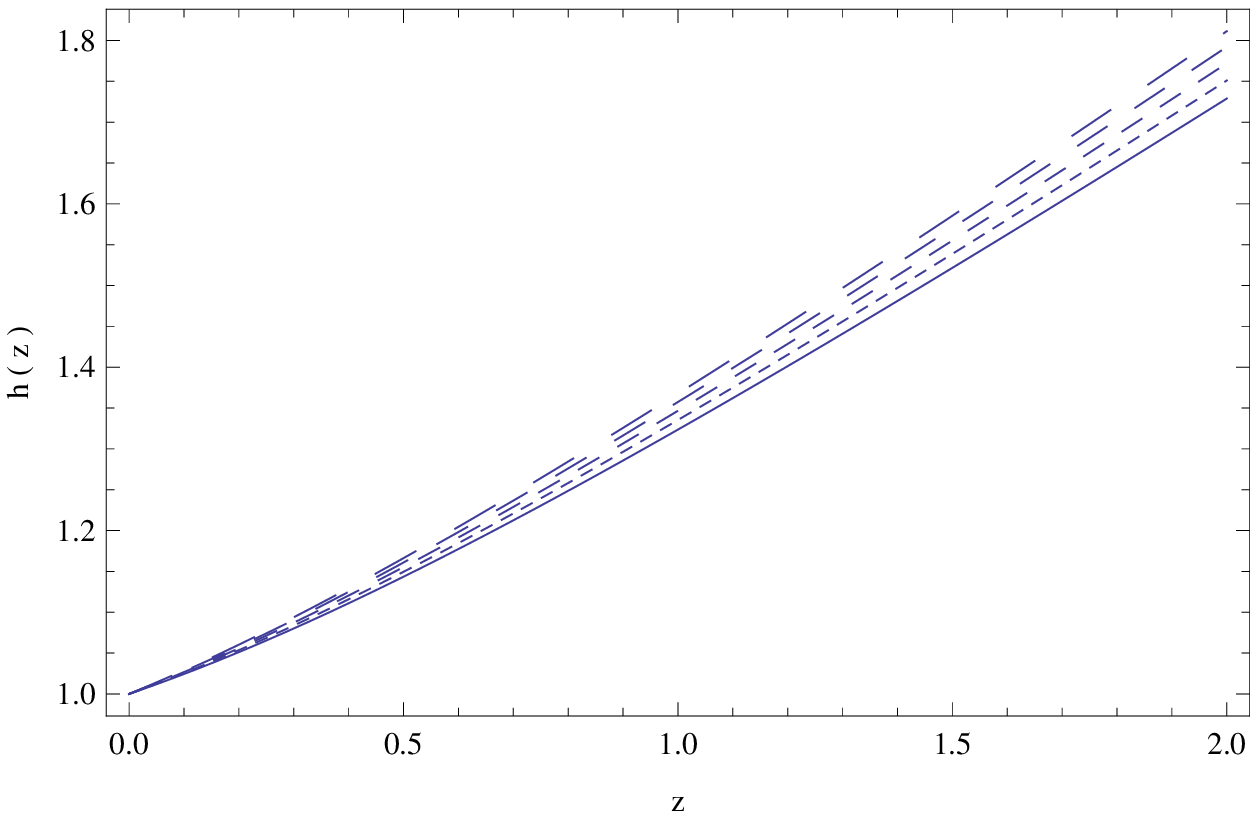}
\end{minipage}
\begin{minipage}{0.5\textwidth}
\includegraphics[width=7.5cm]{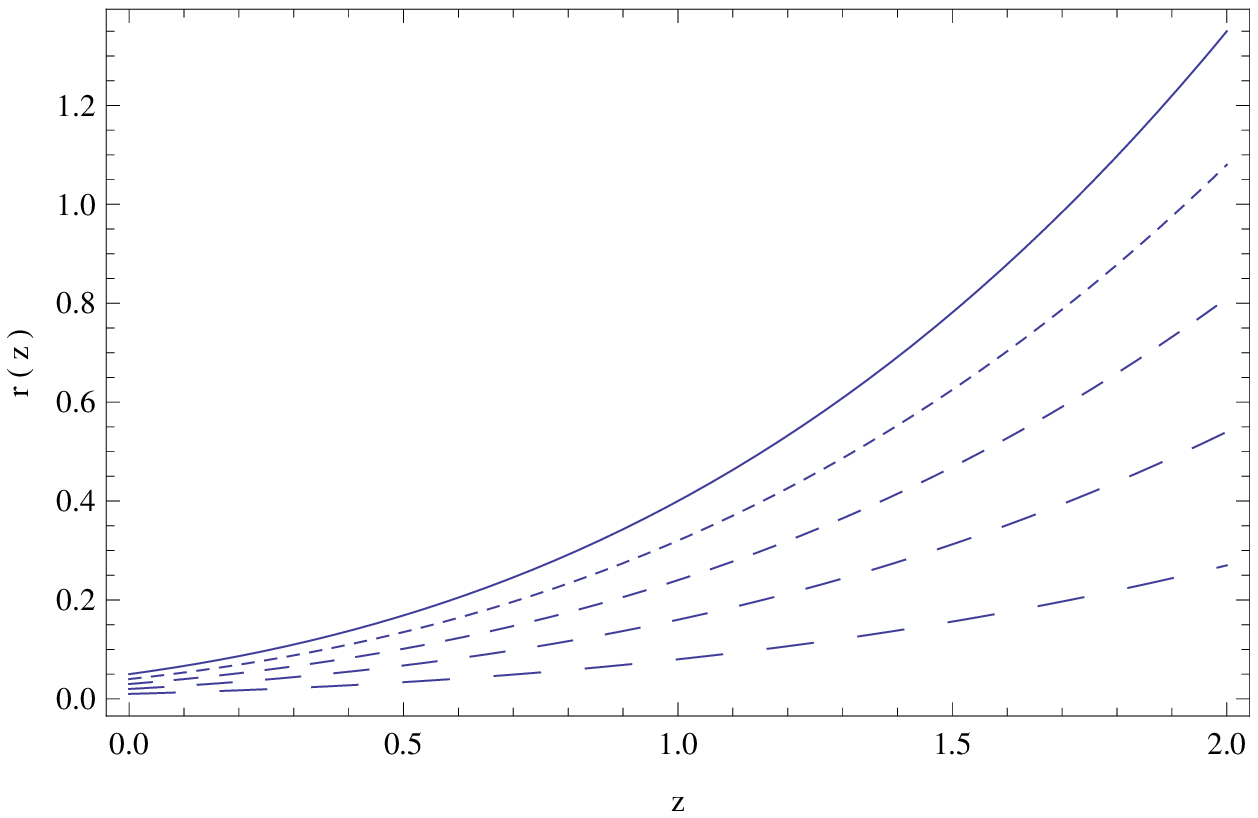}
\end{minipage}
\caption{Variation as a function of $z$ of the Hubble parameter $h(z)$ (left
figure) and of the matter energy density $r(z)$ (right figure) for a
dust universe, for $m=0.002$, $s=0.20$, and for different values of the
parameters $n$:
 $n = 1.653$ (solid curve),  $n = 1.663$ (dotted curve),  $n =
1.673$ (short dashed curve),  $n = 1.683$ (dashed curve), and $n = 1.693$,
respectively.}
\label{fig7_0}
\end{figure}

\begin{figure}[h]
\centering
\includegraphics[width=8cm]{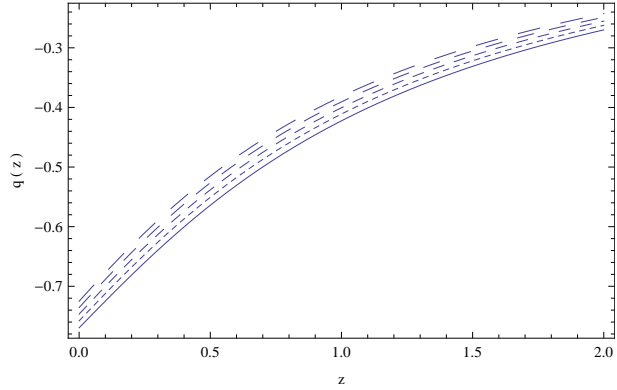} %
\caption{Variation as a function of $z$ of the deceleration parameter $q(z)$
for
a
dust universe, for $m=0.002$, $s=0.20$, and for different values of the
parameters $n$:
 $n = 1.653$ (solid curve),  $n = 1.663$ (dotted curve),  $n =1.673$ (short
dashed curve),  $n = 1.683$ (dashed curve), and $n = 1.693$,
respectively.}
\label{fig8_0}
\end{figure}

The variations with respect to the redshift $z$ of the Hubble function, matter
energy density
and deceleration parameter for the dust universe are represented in
Figs.~\ref{fig7_0} and \ref{fig8_0}.

Both the Hubble function and the matter energy density, shown in
Figs.~\ref{fig7_0}, are monotonically decreasing functions of time. The
deceleration parameter, represented in Fig.~\ref{fig8_0} has negative values in
the range $0\leq z\leq 2$, showing that for the given numerical values of
parameters $m$, $n$ and $s$, the universe experiences an accelerated expansion.

\section{The special matter-aether coupling $\phi T$}\label{sec4}

In this Section we consider a special choice for the coupling between matter and
the aether field. This coupling is common in massive gravity theories and
galileons theories. In order to do this, we add a $\alpha_6 \phi T$ term to the
action %
Eq.~\eqref{1}, where $\alpha_6 $ is a coupling constant, and $T$ is the trace of
the matter energy-momentum tensor. One can easily see that this new coupling
will add a new term  $\alpha_6 T$ to the left hand side of the aether equation 
of
motion, %
Eq.~\eqref{7}, and a term of the form $A_{\mu\nu}$ to the right hand side of
the
metric field
equation Eq.~\eqref{4}, where
\begin{align}  \label{24}
A_{\mu\nu}=\alpha_6\left(\phi T_{\mu\nu}+\frac{1}{2}(T-2\mathcal{L}_m)\phi
g_{\mu\nu}+2\phi g^{\alpha\beta}\frac{\partial^2 \mathcal{L}_m}{\partial
g^{\mu\nu}g^{\alpha\beta}}\right),
\end{align}
with $\mathcal{L}_m$ as the matter Lagrangian (For more detailed 
calculations, see \cite{fRT}).

We note that the energy-momentum tensor of the ordinary matter is not
conserved due to the non-minimal interaction between scalar field and the
energy-momentum tensor. In this case, the non-conservation equation can be
obtained easily by taking the covariant divergence of equation \eqref{4} and
using equation \eqref{7}, with the result
\begin{align}\label{24.1}
\nabla^\mu
T_{\mu\nu}=\f{\alpha_6}{1+\alpha_6\phi}\left[\phi^\mu(g_{\mu\nu}\mathcal{L}_m-T_
{\mu\nu}
)+\phi\nabla_{\nu}(\mathcal{L}_m-\f12 T)-2\nabla^\mu(\phi B_{\mu\nu})\right],
\end{align}
where we have defined $B_{\mu\nu}=g^{\alpha\beta}\f{\partial^2 
\mathcal{L}_m}{\partial
g^{\mu\nu}\partial g^{\alpha\beta}}$. The non-conservation of the
energy-momentum tensor implies that the point particle does not follow the
geodesic equation. In order to obtain the equation of motion for a point
particle, we take assume that the energy-momentum tensor is described by a
pressure-less perfect fluid, with
\begin{align}\label{24.2}
T_{\mu\nu}=\rho u_\mu u_\nu.
\end{align}
Substituting this into \eqref{24.1} and using the relation
\begin{align}\label{24.3}
h^{\nu\lambda}\nabla^\mu T_{\mu\nu}=u^\mu\nabla_\mu
u^\lambda=\f{d^2x^\lambda}{ds^2}+\Gamma^\lambda_{~\rho\sigma}\f{dx^\rho}{ds}\f{
dx^\sigma}{ds},
\end{align}
we obtain the equation of motion as
\begin{align}\label{24.4}
\f{d^2x^\lambda}{ds^2}+\Gamma^\lambda_{~\rho\sigma}\f{dx^\rho}{ds}\f{
dx^\sigma}{ds}=-\f{h^{\nu\lambda}}{\rho}\f{\alpha_6}{1+\alpha_6\phi}\left[
\f12\phi\nabla_\nu\rho+\rho\phi_\nu\right].
\end{align}
In the case $\alpha_6=0$ we obtain the standard geodesic equation.
\subsection{The cosmological evolution in the presence of the $\phi T$ coupling}

In order to obtain the effect of this new coupling term on the cosmological
evolution of the universe, we note
that one can choose the Lagrangian for a perfect fluid as $\mathcal{L}_m=-\rho$. 
Then for
a flat Friedmann-Robertson-Walker geometry we obtain
the metric and aether equations as
\begin{align}
&\frac{3}{2}(15\alpha_4+2\kappa^2)H^2-21\alpha_5 H^3-3\alpha_3
H-\rho+\lambda+\f{\alpha_6}{2}(3p-\rho)t=0,  \label{25.1} \\
&(3\alpha_4+2\kappa^2-6\alpha_5H)\dot{H}+\frac{3}{2}
(3\alpha_4+2\kappa^2)H^2-6\alpha_5H^3+p+\f{\alpha_6}{2}(5p+\rho)t=0 ,
\label{25.2}
\end{align}
and
\begin{align}  \label{26}
6(15\alpha_5H^2-12\alpha_4H+\alpha_3)\dot{H}
+90\alpha_5H^4-108\alpha_4H^3+18\alpha_3H^2-2(3\lambda
H+\dot{\lambda})-\alpha_6(3p-\rho)=0,
\end{align}
where we assume that $c_1=0$ in Eq.~\eqref{cos-2}.

As a first step in
the analysis of the system of equations Eqs.~(\ref{25.1}) - (\ref{26}) we
rescale again the physical parameters according to
\begin{align}
&\alpha _3=\frac{2\kappa ^2}{3}H_0\eta, ~\alpha _4=\frac{2\kappa
^2}{15}\sigma,~\alpha _5=\frac{2\kappa ^2}{21H_0}%
\theta ,~H=H_0h,\nonumber\\
&\rho =2\kappa ^2H_0^2r,~p=2\kappa ^2H_0^2P,~ \lambda
=2\kappa
^2H_0^2\Lambda,~\alpha_6 =H_0\Delta, ~t=\frac{\tau }{H_0},
\end{align}
where $H_0$ is the present day value of the Hubble parameter and $\eta $,
$\sigma $ and $\theta $
are dimensionless constants. Then the system of Eqs.~(\ref{25.1}) - (\ref{26}%
) takes the following dimensionless form
\begin{align}  \label{f1}
\frac{3}{2}(1+\sigma)h^2(\tau)-\theta h^3(\tau)-\eta h(\tau)-r(\tau)+\Lambda
(\tau)+\frac{\Delta }{2}\left[3P(\tau )-r(\tau )\right]\tau =0,
\end{align}
\begin{align}  \label{f2}
\left[1+\frac{\sigma }{5}-\frac{2}{7}\theta h(\tau)\right]\frac{dh(\tau
)}{d\tau}+\frac{3}{2}\left(1+\frac{\sigma
}{5}\right)h^2(\tau)-\frac{2}{7}\theta
h^3(\tau )+P+\frac{\Delta }{2}\left[5P(\tau )+r(\tau)\right]\tau =0,
\end{align}
\begin{align}  \label{f3}
&6\left[\frac{5}{7}\theta h^2(\tau)-\frac{4}{5}\sigma h(\tau)+\frac{\eta
}{3}\right]\frac{dh(\tau)}{d\tau }+\frac{30}{7}\theta h^4(\tau)
-\frac{36}{5}\sigma h^3(\tau)\nonumber\\&\qquad\qquad\qquad+6\eta h^2(\tau)
-2\left[3\Lambda (\tau)h(\tau )+\frac{d\Lambda (\tau )}{d\tau }\right]-\Delta
\left[3P(\tau )-r(\tau )\right] =0.
\end{align}
The above equations can not be solved analytically. Hence in the following
we consider their numerical solutions for two particular choices of the
equation of state of the cosmological matter.

\subsection{The radiation fluid cosmological model}

In the case of a universe filled with a radiation fluid, $P(\tau )=r(\tau)/3$,
the
system of Eqs.~(\ref{f1}) - (\ref{f3}) takes the form
\begin{align}
r(\tau )=\frac{3}{2}(1+\sigma)h^2(\tau)-\theta h^3(\tau )-\eta h(\tau)+\Lambda
(\tau),
\end{align}%
\begin{align}  \label{68}
\frac{dh\left( \tau \right) }{d\tau }&=\frac{1}{21(\sigma +5)-30\theta h(\tau
)}\Bigg[-35(4\Delta \tau +1)\Lambda
(\tau )+35h(\tau )(4\Delta \eta \tau +\eta )\nonumber\\&\qquad+5\theta
(28\Delta \tau
+13)h^3(\tau )-42h^2(\tau )\left[ 5\Delta (\sigma +1)\tau +2\sigma +5%
\right]\Bigg],
\end{align}%
\begin{align}  \label{69}
\frac{d\Lambda \left( \tau \right) }{d\tau }&=\frac{1}{105\left[ 7(\sigma
+5)-10\theta h(\tau )\right] }\Bigg\{ 35(4\Delta \tau +1)\Lambda (\tau
) \left[ -35\eta -75\theta h^2(\tau
)+84\sigma h(\tau )\right]\nonumber\\
&-35h(\tau )(4\Delta \eta \tau +\eta )-5\theta (28\Delta \tau +13)h^3(\tau
)+42h^2(\tau )\left[ 5\Delta (\sigma +1)\tau +2\sigma +5\right]
\Bigg\}\nonumber\\
&+3\eta h^2(\tau )+\frac{15}{7}\theta h^4(\tau )-3h(\tau
)\Lambda (\tau )-\frac{18}{5}\sigma h^3(\tau ).
\end{align}

\begin{figure}[h]
\begin{minipage}{0.5\textwidth}
\includegraphics[width=7.5cm]{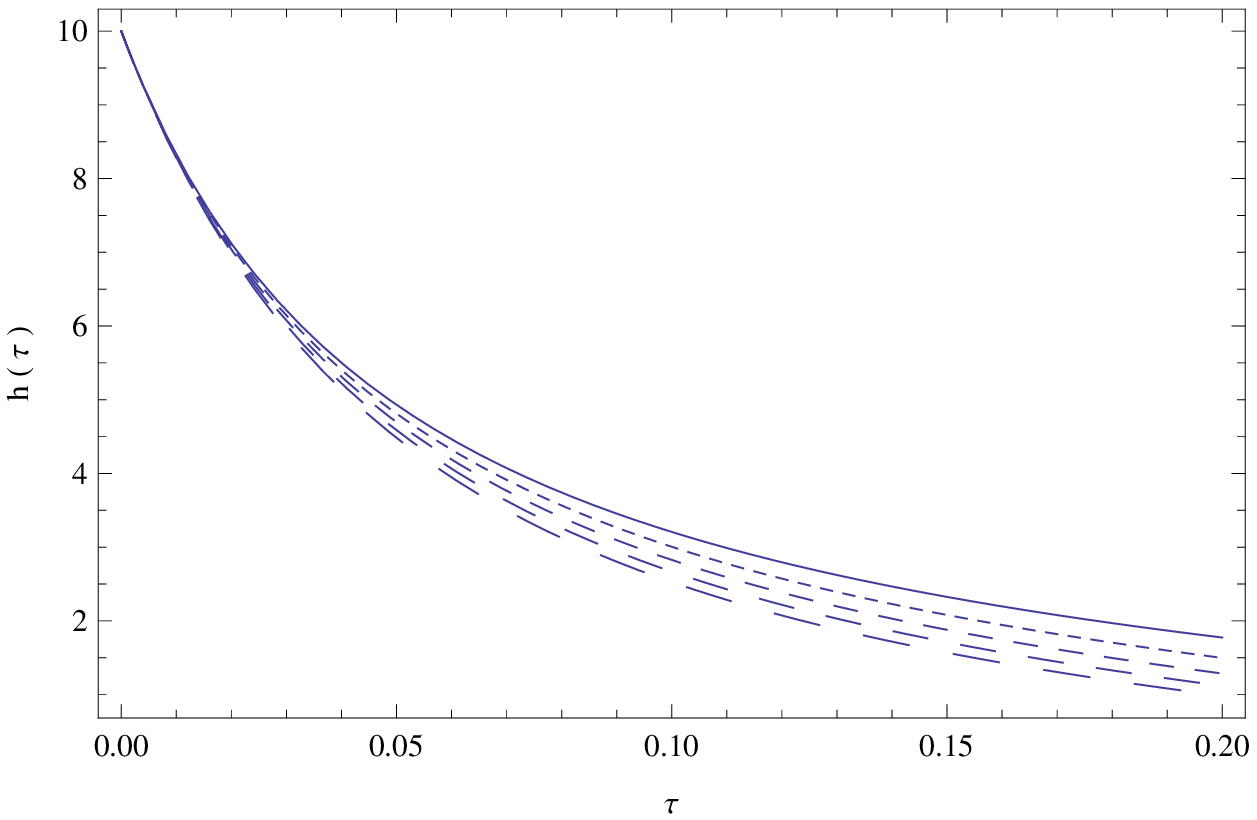}
\end{minipage}
\begin{minipage}{0.5\textwidth}
\includegraphics[width=7.5cm]{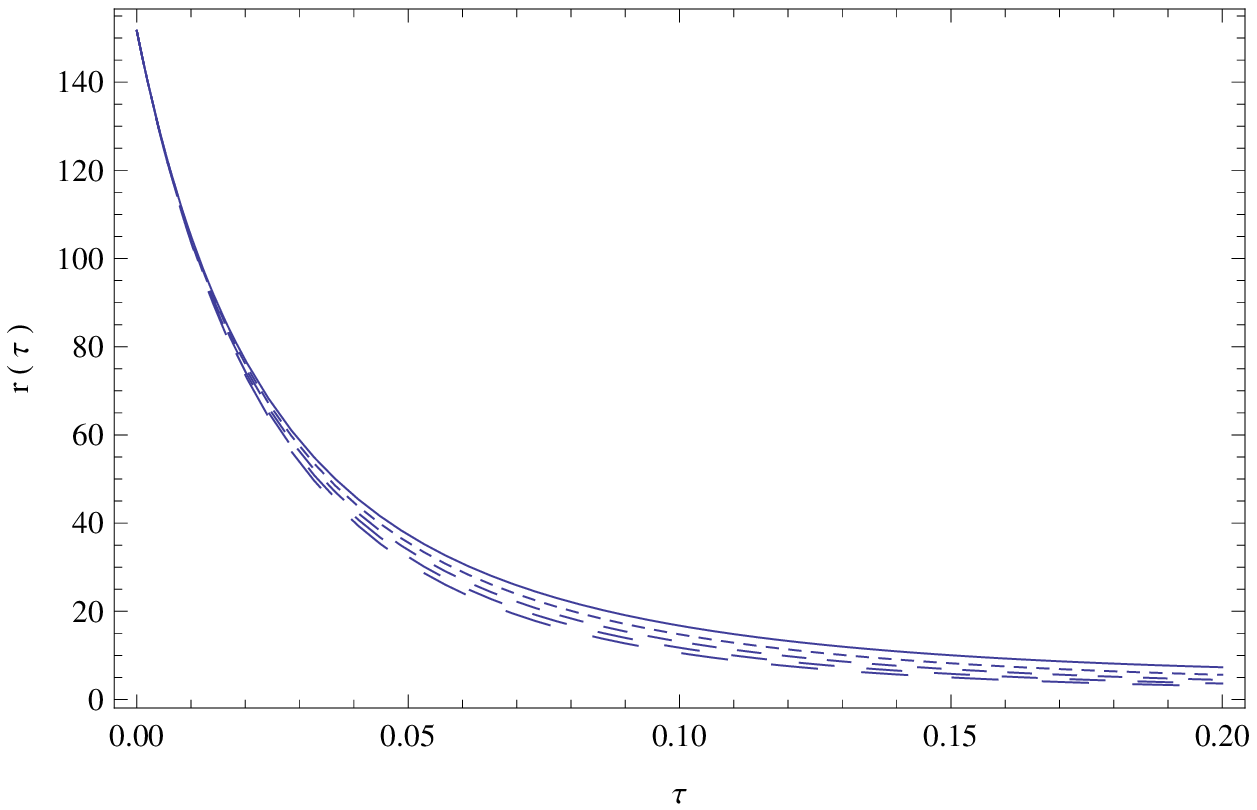}
\end{minipage}
\caption{Variation with respect to $\tau $ of the Hubble parameter $%
h(\tau)$ (left figure) and of the energy density $r(\tau)$ (right figure) for a
radiation fluid filled universe, for different values
of the parameters $\Delta$: $\Delta = 1$ (solid curve), $\Delta = 3$,
(dotted curve), $\Delta = 5$ (short dashed curve), $\Delta = 7$ (dashed
curve), and $\Delta = 9$ (long dashed curve), respectively.  }
\label{fig7}
\end{figure}

\begin{figure}[h]
\begin{minipage}{0.5\textwidth}
\includegraphics[width=7.5cm]{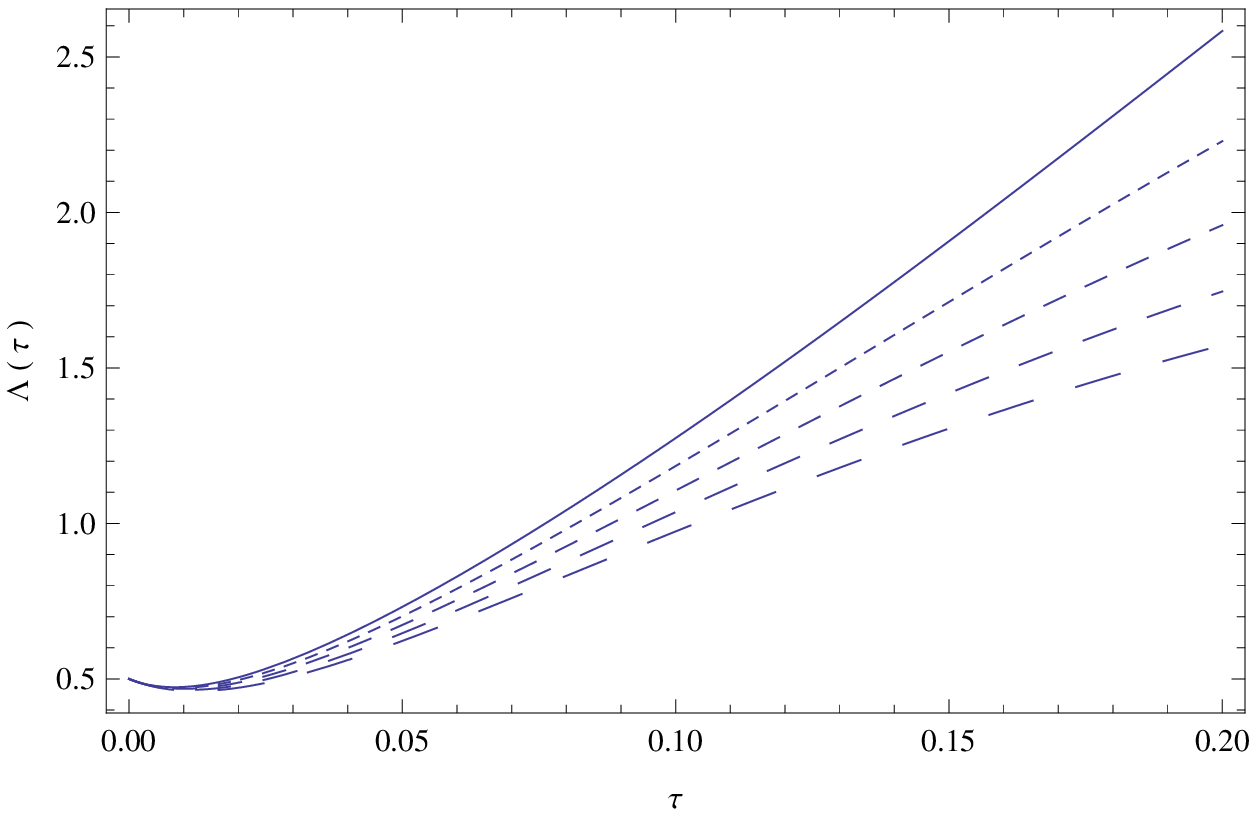}
\end{minipage}
\begin{minipage}{0.5\textwidth}
\includegraphics[width=7.5cm]{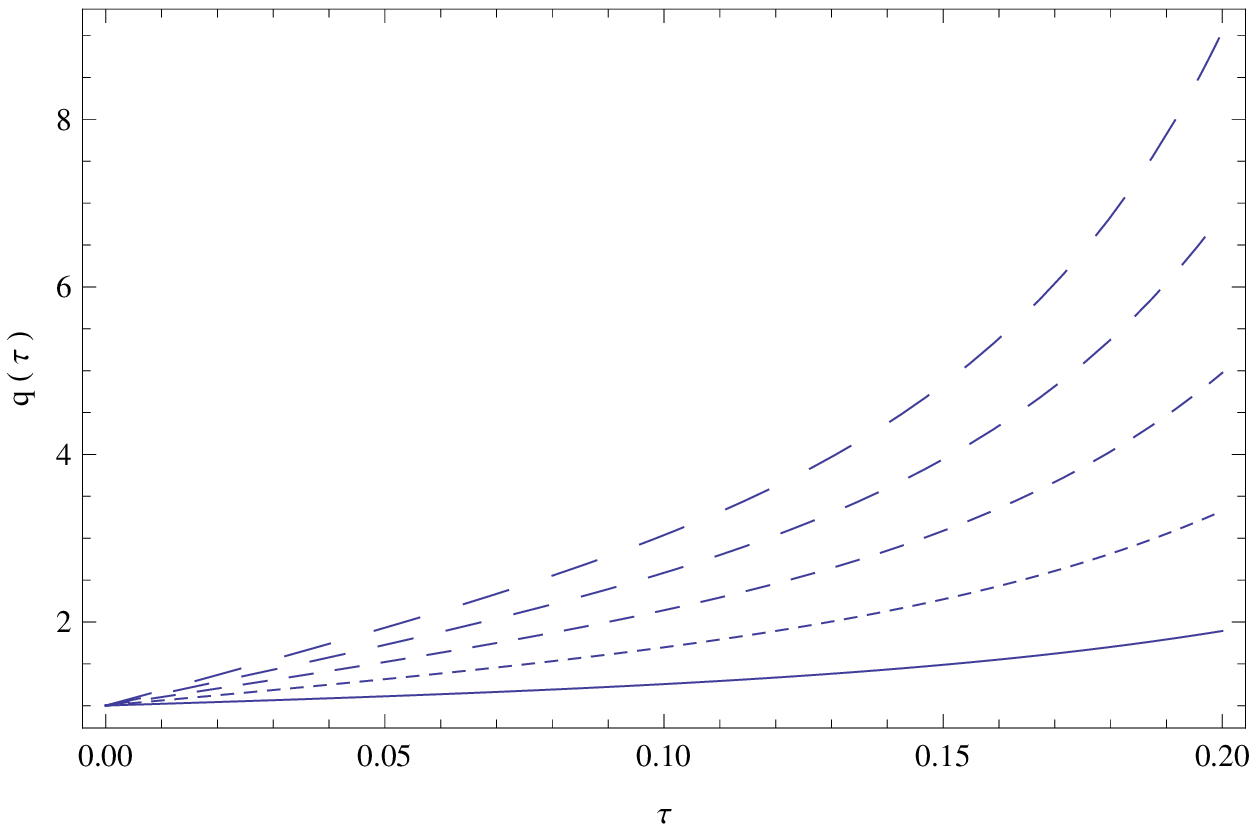}
\end{minipage}
\caption{Variation with respect to $\tau $ of the Lagrange
multiplier $\Lambda (\tau)$ (left figure) and of the deceleration parameter
$q(\tau)$ (right figure) for a radiation fluid filled universe,
for different values of the parameters $\Delta$: $\Delta = 1$ (solid curve),
$\Delta = 3$,
(dotted curve), $\Delta = 5$ (short dashed curve), $\Delta = 7$ (dashed
curve), and $\Delta = 9$ (long dashed curve), respectively.  }
\label{fig8}
\end{figure}

The deceleration parameter can be obtained as
\begin{align}
q(\tau )&=\frac{1}{3 h^2(\tau ) (10 \theta  h(\tau )-7 (\sigma
   +5))}\Bigg[7 h(\tau ) \bigg(5 (4 \Delta  \eta  \tau +\eta )+5 h^2(\tau ) (4
\Delta  \theta
    \tau +\theta )\nonumber\\&\qquad-3 h(\tau ) (10 \Delta  (\sigma +1) \tau +3
\sigma +5)\bigg)-35 (4
   \Delta  \tau +1) \Lambda (\tau )\Bigg].
\end{align}
The time variations of the Hubble parameter, of the energy density, of the
Lagrange multiplier,
 and of the deceleration parameter of the radiation fluid
filled universe are presented, for different values of the parameter $\Delta $,

in Figs.~\ref{fig7} and \ref{fig8}. In all cases $\sigma =0.001$, $\theta
=-0.001$, and $\eta =0.001$.
 The initial conditions used to numerically integrate Eqs.~(%
\ref{68}) and (\ref{69}) are $h(0)=10$ and $\Lambda (0)=0.5$.

In the plots we have adopted constant values for the parameters $\sigma$,
$\theta $ and $\eta$,
and we have varied the numerical value of $\Delta $, describing the strength
of the coupling between $\phi $ and $T$.  As one can see from Fig.~(\ref%
{fig7}), the Hubble parameter is a monotonically decreasing function of
time. The Lagrange multiplier $\Lambda $, presented in Fig.~(\ref{fig8}),
has a complex cosmological evolution, consisting of two phases: a short, almost
linearly decreasing with time which ends with reaching a minimum value at a
finite
time followed by a monotonic increase phase. The energy density $r$, shown in
Fig.~(\ref{fig7}), is a monotonically decreasing function of time. As seen
from Fig.~\ref{8}, the dynamics of the deceleration parameter indicates
that for all the chosen numerical values of the model parameters the radiation
filled universe
 is in a decelerating phase. The time spent in the decelerating
phase strongly depends on the numerical values of $\Delta $.

\subsection{The dust cosmological model}

\begin{figure}[h]
\begin{minipage}{0.5\textwidth}
\includegraphics[width=7.5cm]{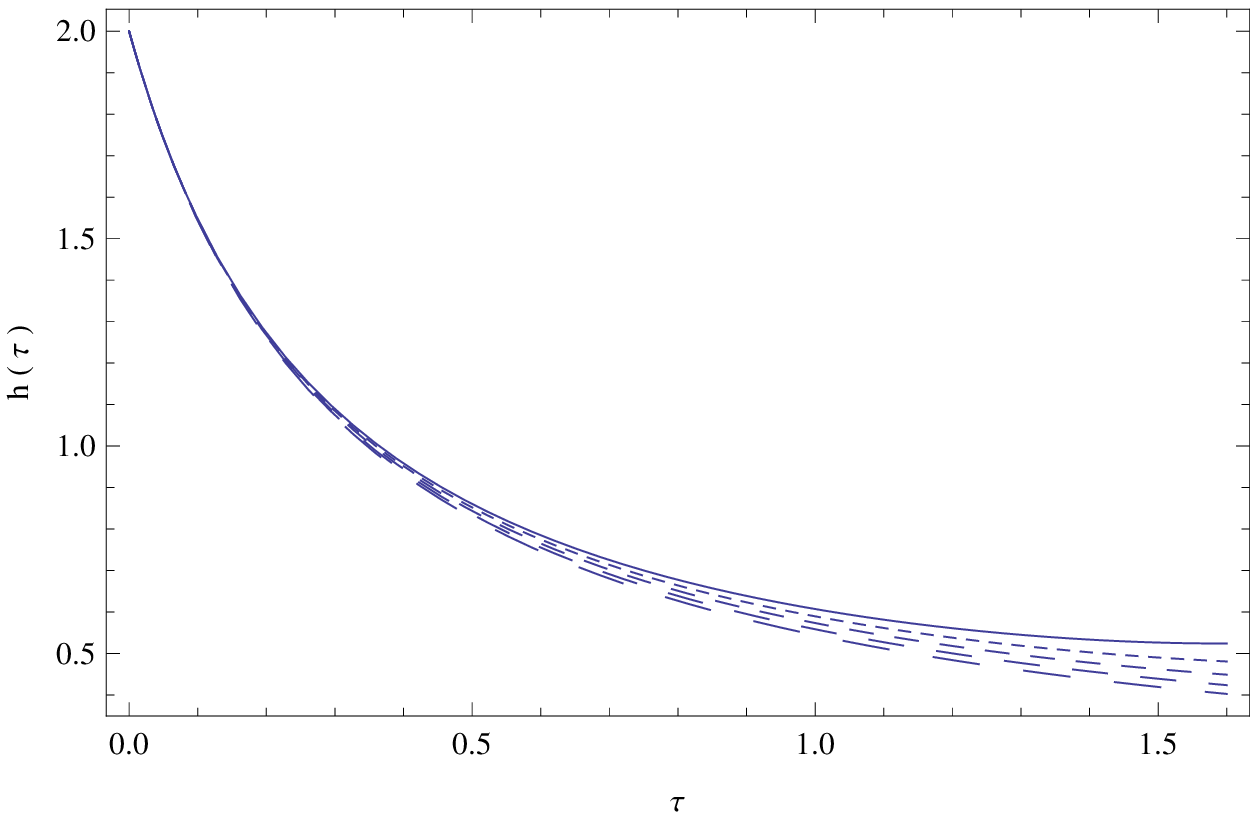}
\end{minipage}
\begin{minipage}{0.5\textwidth}
\includegraphics[width=7.5cm]{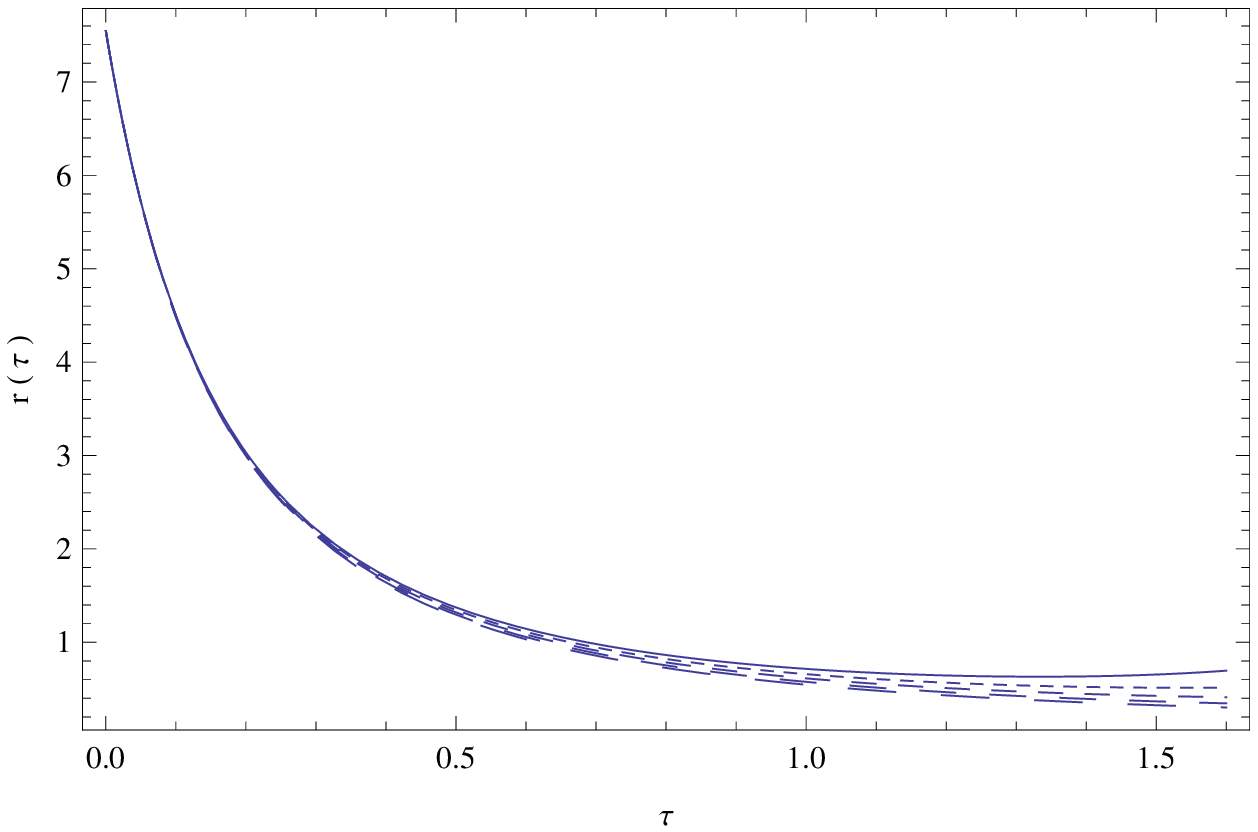}
\end{minipage}
\caption{Variation with respect to the dimensionless time $\tau $ of
the Hubble parameter $h(\tau)$ (left figure) and of the matter energy density
$r(\tau)$ (right figure),  for a dust universe, for different
values of the parameters $\Delta$: $\Delta = -0.75$ (solid curve), $\Delta =
-0.65$, (dotted curve), $\Delta = -0.55$ (short dashed curve), $\Delta = -0.45$
(dashed curve), and $\Delta = -0.35$ (long dashed curve), respectively. }
\label{fig11}
\end{figure}

As a second example of a cosmological model in the framework of the SEA
theory with aether-matter coupling we consider the case of a universe filled
with dust, that is, matter with zero pressure. By taking $P=0$ the
cosmological equations describing the evolutionary dynamics of the universe
become
\begin{align}
r(\tau )=\frac{-\eta  h(\tau )-\theta  h^3(\tau )+(3/2) (\sigma +1) h^2(\tau
)+\Lambda
   (\tau )}{\Delta  \tau /2+1},
\end{align}%
\begin{align}  \label{f3_1}
\frac{dh(\tau )}{d\tau }=\frac{-35\Delta \tau \Lambda (\tau )+35\Delta \eta
\tau h(\tau )+5\theta (9\Delta \tau +4)h^{3}(\tau )-21h^{2}(\tau )\left[
\Delta (3\sigma +5)\tau +\sigma +5\right] }{(\Delta \tau +2)\left[ 7(\sigma
+5)-10\theta h(\tau )\right] },
\end{align}
\begin{align}  \label{f4}
\frac{d\Lambda \left( \tau \right) }{d\tau }&=\frac{1}{2(\Delta \tau +2)\left[
7(\sigma +5)-10\theta h(\tau )\right] }\Big[ 14\Delta \Lambda (\tau
)(-5\eta \tau +\sigma +5)\nonumber\\&+h^{2}(\tau )(\Delta (4\eta (5\theta
-63\sigma \tau
)+21(\sigma +1)(\sigma +5))+30\theta (4-3\Delta \tau )\Lambda (\tau )\nonumber\\
&+42\eta
(\sigma +5))+2h(\tau )(7\Delta \eta (5\eta \tau -\sigma -5)-\Lambda (\tau
)(\Delta (10\theta +21(5-3\sigma )\tau )+42(\sigma +5)))\nonumber\\
&+4h^{3}(\tau
)(9\Delta \tau (5\eta \theta +7\sigma (\sigma +1))-\Delta \theta (11\sigma
+25)-20\eta \theta )+150\Delta \theta ^{2}\tau h^{5}(\tau )\nonumber\\&+2\theta
h^{4}(\tau )(10\Delta \theta -6\Delta (32\sigma +25)\tau +9\sigma
-75)\Big].
\end{align}
The deceleration parameter can be obtained as
\begin{align}
q(\tau)=\frac{7 h(\tau ) (5 \Delta  \eta  \tau -h(\tau ) (2 \Delta  (4 \sigma
+5) \tau -5 \Delta  \theta  \tau  h(\tau )+\sigma +5))-35 \Delta  \tau  \Lambda
(\tau )}{(\Delta  \tau +2)
   h^2(\tau ) [10 \theta  h(\tau )-7 (\sigma +5)]}.
\end{align}
The time variation of the Hubble parameter $h(\tau)$, of the energy density
$r(\tau)$, of the Lagrange
multiplier $\Lambda (\tau )$  and of the
deceleration parameter $q(\tau)$ are presented in Figs.~\ref{fig11} and \ref%
{fig12}.

\begin{figure}[h]
\begin{minipage}{0.5\textwidth}
\includegraphics[width=7.5cm]{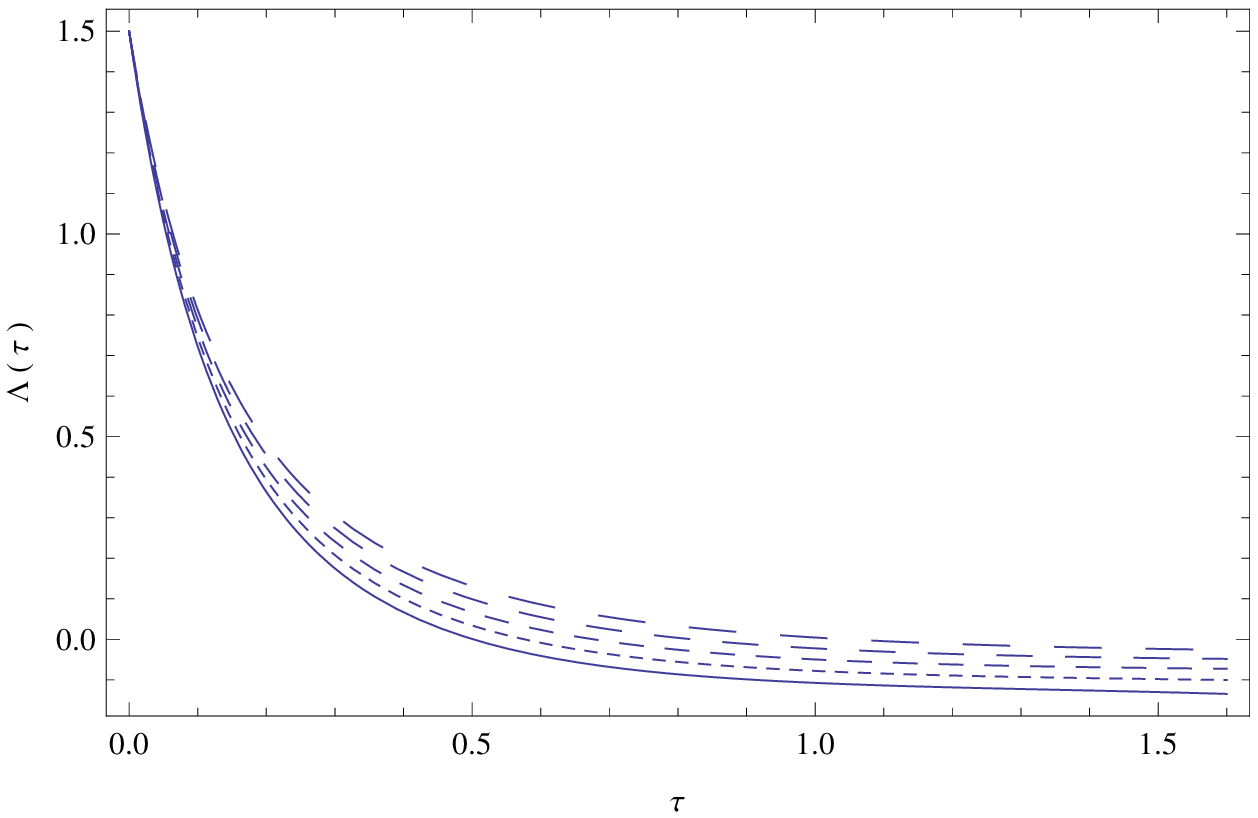}
\end{minipage}
\begin{minipage}{0.5\textwidth}
\includegraphics[width=7.5cm]{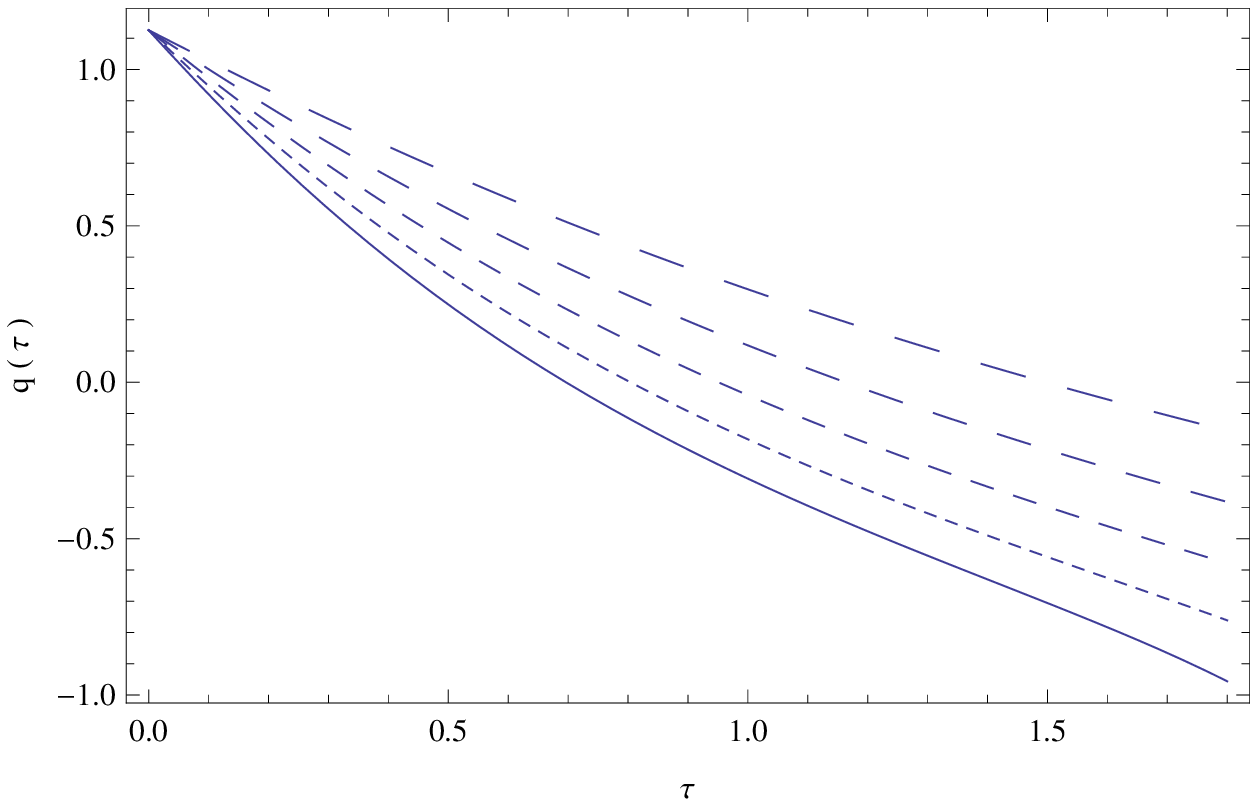}
\end{minipage}
\caption{Variation with respect to the dimensionless time $\tau $ of
the Lagrange multiplier $\Lambda (\tau)$ (left figure) and of the deceleration
parameter $q(\tau)$ (right figure) for a dust universe, for
different values of the parameters $\Delta$: $\Delta = -0.75$ (solid curve),
$\Delta =
-0.65$, (dotted curve), $\Delta = -0.55$ (short dashed curve), $\Delta = -0.45$
(dashed curve), and $\Delta = -0.35$ (long dashed curve), respectively. }
\label{fig12}
\end{figure}

Here, in a similar fashion to that of the radiation fluid filled universe model,
we have also fixed the numerical values of the free parameters $\sigma$,
$\theta$  and $\eta $ as $\sigma =0.001$, $\theta =-0.005$, $\eta =0.001$,
and have varied the coupling parameter $\Delta <0$. The initial conditions used
to integrate the field equations Eqs~(\ref{f3_1}) and (\ref{f4}) are $h(0)=2$
and
$\Lambda (0)=1.5$, respectively. The Hubble parameter,
depicted in Fig.~\ref{fig11}, is a monotonically decreasing function of
time, as well as  the energy density, shown in Fig.~\ref{fig11}. The Lagrange
multiplier $\Lambda $, plotted in Fig.~\ref{fig12}, shows
a similar behavior as for the radiation fluid cosmological model,
decreasing monotonically with time. The deceleration parameter,
depicted in Fig.~\ref{fig12}, starts its evolution at a high redshift (high
value of $h$) with a positive value of the order of $q\approx 1$, indicating an
initially decelerating cosmological evolution. Due to the presence of the
aether field and of its coupling with matter the deceleration parameter
monotonically decreases in time and the dust filled universe enters  an
accelerating stage, with the deceleration parameter reaching values of the
order
of $q\approx -1$.

\section{Discussion and final remarks}
In this paper we have studied a Lorentz violating theory of gravity by
introducing a time-like vector field in the gravitational action. The time-like
vector field  is constructed from the convariant derivative of the
scalar field $\phi$. In order to impose the time-like property of the vector
$\nabla_\mu\phi$, we have added the constraint
$\nabla_\mu\phi\nabla^\mu\phi+1=0$ to the action through a Lagrange multiplier.
In order to add a kinetic term for the scalar field $\phi$, we note that
the canonical kinetic term for the scalar field is already used in the
constraint equation. With this constraint term, the scalar field becomes dynamical in the theory \cite{cauchy}. In order to enrich the dynamics of the scalar field one may add some higher order
derivative scalar interaction terms to the action. This was done in \cite{SEA}
where the authors used the most general kinetic terms of a vector field to
construct such  interaction terms. However, the resulting theory suffers from
ghost instabilities because of these higher order derivative interactions. This 
can be seen by noting that the theory \cite{SEA} is equivalent to the 
projectable version of Horava-Lifshitz gravity which suffers from instabilities 
and strong coupling in low energies \cite{covHL,SEA,HL}. In
this paper we have added some special higher order derivative interactions
which produce second order field equation, and hence, no Ostrogradski
instabilities. These interactions are known as Galileon terms \cite{gali}.	

The present model can also be considered as a generalization of the mimetic
dark matter models \cite{14}. In the mimetic dark matter model, one defines a
physical metric out of a scalar field and some primary metric tensor in such a
way that the physical metric becomes conformally invariant. Thus constructing a
gravitational theory by the physical metric will automatically produce a
conformally invariant gravitational theory. However, as was mentioned in
\cite{15}, this theory is equivalent to a gravitational theory coupled to a
scalar field and subject to the constraint that $\nabla_\mu\phi$ is time-like.
This is in fact equivalent to the scalar Einstein-aether theory \cite{SEA}. In this
sense, our present model is a generalization of the mimetic dark matter theory
with some additional healthy derivative self-interaction terms. This will
actually make the dark matter imperfect \cite{17}. The problem of whether this
model can satisfy  dark matter experimental data will be considered elsewhere.

In the present paper, as a first step in the in-depth investigation of the
model, we have studied the cosmological consequences of the Lorentz
violating Galileon theory. If one assumes that the matter content of the
universe consists of radiation or stiff matter, then the theory predicts a decelerating universe. In
this case the universe decelerates more rapidly with stiff matter
as compared to  radiation. For a dust baryonic matter the model
suggests an  accelerating universe with an exponential de Sitter acceleration
at late times.

We have also considered a special coupling between ordinary matter and the
scalar field. This coupling will break the conservation law of the ordinary
matter and hence one can expect some modification in the predictions of the
Solar System tests, which may put a constraint on the value of the coupling constant
$\alpha_6$ in Eq.~\eqref{24}. This will also be done in a separate work. In this
paper, we have considered only the cosmological implications of this extra
term. In
the radiation or stiff matter cases, the quantitative behavior of the universe
does not change as compared to previous models without the interaction term,
predicting a decelerating universe. In the case of a dust baryonic matter, however, one can obtain a
universe starting from a decelerating phase and entering into an
accelerating phase at later times.
Other astrophysical and theoretical implications of this model such as black
hole solutions and the stability of the cosmological solutions will be studied in a future publication.

\section*{Acknowledgments}

 T. H. would like to thank the Department of Physics of the Sun-Yat Sen
University in Guangzhou, P. R. China,  for the kind hospitality offered during
the preparation of this work.

\end{document}